%% file: main.tex
\documentclass[twocolumn,twocolappendix]{aastex63}
\usepackage{lineno}

\newcommand{\madcows}{MaDCoWS}
\newcommand{\chandra}{{\it Chandra}}

\newcommand{\uat}[2]{\href{http://astrothesaurus.org/uat/#2}{#1 (#2)}}
\defcitealias{lehmer_2012}{L12}
\defcitealias{nurgalive_2013}{N13}
\defcitealias{nurgaliev_2017}{N17}

\shorttitle{AGNs in Dynamically Active \madcows\ Clusters}
\shortauthors{Muhibullah et al.}

\usepackage{mathtools}
\usepackage{enumitem}
\setlist{nolistsep}

\received{April 30, 2024}
\revised{January 28, 2025}
\accepted{February 24, 2025}
\submitjournal{ApJ}

\begin{document}

\title{\large The Massive and Distant Clusters of \textit{WISE} Survey. XII. Exploring X-ray AGN in Dynamically Active Massive Galaxy Clusters at $z\sim1$}

\correspondingauthor{Mustafa Muhibullah}
\email{mmuhibullah@crimson.ua.edu}
\author[0000-0002-0347-0900]{Mustafa Muhibullah} 
\affiliation{Department of Physics and Astronomy, The University of Alabama, Tuscaloosa, AL 35487, USA}
\affiliation{Department of Physics and Astronomy, University of Missouri, 5110 Rockhill Road, Kansas City, MO, 64110, USA}

\author[0000-0002-4208-798X]{Mark Brodwin}
\affiliation{Department of Physics and Astronomy, University of Missouri, 5110 Rockhill Road, Kansas City, MO, 64110, USA}

\author[0000-0001-5226-8349]{Michael McDonald}
\affiliation{Kavli Institute for Astrophysics and Space Research, Massachusetts Institute of Technology, 70 Vassar St, Cambridge, MA 02139, USA}

\author[0000-0002-0933-8601]{Anthony H. Gonzalez}
\affiliation{Department of Astronomy, University of Florida, 211 Bryant Space Center, Gainesville, FL 32611, USA}

\author[0000-0001-9793-5416]{Emily Moravec}
\affiliation{Green Bank Observatory, P.O. Box 2, Green Bank, WV 24944, USA}

\author[0000-0002-7898-7664]{Thomas Connor}
\affiliation{Center for Astrophysics $\vert$\ Harvard\ \&\ Smithsonian, 60 Garden St., Cambridge, MA 02138, USA}
\affiliation{Jet Propulsion Laboratory, California Institute of Technology, 4800 Oak Grove Drive, Pasadena, CA 91109, USA}
 
\author[0000-0003-0122-0841]{S. A. Stanford}
\affiliation{Department of Physics, University of California, One Shields Avenue, Davis, CA 95616, USA}

\author[0000-0002-0955-8954]{Florian Ruppin}
\affiliation{Univ. Lyon, Univ. Claude Bernard Lyon 1, CNRS/IN2P3, IP2I Lyon, F-69622, Villeurbanne, France}

\author[0000-0003-3521-3631]{Taweewat Somboonpanyakul}
\affiliation{Department of Physics, Faculty of Science, Chulalongkorn University 254 Phayathai Road, Pathumwan, Bangkok 10330, Thailand}

\author{Peter R. M. Eisenhardt}
\affiliation{Jet Propulsion Laboratory, California Institute of Technology, 4800 Oak Grove Drive, Pasadena, CA 91109, USA}

\author[0000-0002-5674-5082]{Bandon Decker}
\affiliation{Department of Physics and Astronomy, University of Missouri, 5110 Rockhill Road, Kansas City, MO, 64110, USA}

\author[0000-0003-2686-9241]{Daniel Stern}
\affiliation{Jet Propulsion Laboratory, California Institute of Technology, 4800 Oak Grove Drive, Pasadena, CA 91109, USA}

\author[0000-0003-3428-1106]{Ariane Trudeau}
\affiliation{Department of Astronomy, University of Florida, 211 Bryant Space Center, Gainesville, FL 32611, USA}

\begin{abstract}
We present an analysis of the cluster X-ray morphology and active galactic nucleus (AGN) activity in nine $z\sim1$ galaxy clusters from the Massive and Distant Clusters of \textit{WISE} Survey (\madcows) observed with \chandra.~Using photon asymmetry ($A_{\text{phot}}$) to quantify X-ray morphologies, we find evidence that the four most dynamically disturbed clusters are likely to be mergers.~Employing a luminosity cut of $7.6\times10^{42}$ erg/s to identify AGN in the 0.7--7.0 keV, we show that the majority of these clusters host excess AGN compared to the local field.~We use the cumulative number-count ($\log N- \log S$) model to predict AGN incidence in cluster isophotes under this luminosity cut.~Our analysis finds evidence (at $> 2\sigma$) of a positive correlation between AGN surface densities and photon asymmetry, suggesting that a disturbed cluster environment plays a pivotal role in regulating AGN triggering.~Studying AGN incidence in cluster X-ray isophotes equivalent in area to $1.0r_{500}$, we find that the AGN space density inversely scales with cluster mass as $\sim M^{-0.5^{+0.18}_{-0.18}}$ at the 3.18$\sigma$ level. Finally, when we separately explore the cluster mass dependence of excess AGN surface density in disturbed and relaxed clusters, we see tentative evidence that the two morphologically distinct sub-populations exhibit diverging trends, especially near the outskirts, likely due to cluster merger-driven AGN triggering/suppression.
\end{abstract}

\keywords{
\uat{Galaxy environments}{2029};
\uat{Galaxy clusters}{584};
\uat{High-redshift galaxy clusters}{2007};
\uat{Intracluster medium}{858};
\uat{X-ray active galactic nuclei}{2035};
\uat{X-ray astronomy}{1810}
}

\section{Introduction}\label{sec:intro}
The hot dense environment in a cluster can significantly influence the growth and evolution of its member galaxies \citep{Pimbblet_2013, davidzon_2016,darvish_2017}.~Earlier studies have established that clusters residing in the local Universe are rich in passively evolving, red elliptical galaxies, in contrast to the dominant star-forming blue spirals in the field \citep[e.g.,][]{dressler_1980, balogh_2004}. Both low \citep{best_2004,kauffmann_2004,scoville_2013} and intermediate \citep{kodama_2004} redshift cluster observations have demonstrated that the comoving star formation rate and star-forming fractions decrease with cluster density. Evidence reported by e.g.\ \cite{croton_2006,hopkins_2006, hopkins_2012,laigle_2018} supports the idea that the active galactic nucleus (AGN) triggering mechanism plays a key role in regulating the star-formation activity of the host galaxy. 

The physical forces that drive AGN activity, however, remain poorly constrained, while inherent selection biases and host galaxy properties can prove challenging to our understanding, resulting in diverse interpretations \citep[e.g.,][]{strand_2008,lietzen_2009}. Earlier studies found optically selected luminous AGNs are relatively less frequent in the low-redshift clusters when compared to the field \citep[e.g.,][]{dressler_1985, koulouridis_2010, von_linden_2014}. However, optical AGN selection is biased towards unobscured quasars and inefficient in separating AGNs from star-forming galaxies at high redshift \citep[e.g.,][]{dickey_2016, Overzier_2016, balmaverde_2017}. 

X-ray surveys are capable of detecting AGNs with the highest purity \citep[see Fig.~11 from][]{padovani_2016} and density \citep{luo_2017} due to the penetrating nature of the hot X-ray emissions that originate from the accretion disk of the supermassive black hole \citep[e.g.,][]{brandt_2015}. In the central region of the clusters, several studies found an overabundance of X-ray AGN, particularly within 1--2 Mpc \citep{ruderman_2005,gilmour_2009}, while others claimed significant environmental suppression near the virial radius \citep{martini_2009,ehlert_2013,ehlert_2014}. When probed to lower luminosities, some studies also found no evidence of AGN suppression or enhancement at all \citep{haggard_2010,koulouridis_2014}. Multi-wavelength observations, meanwhile, find substantial redshift evolution of AGN fraction as a function of the environment at $z\geqslant1$ \citep{galametz_2009,martini_2013,bufanda_2017}. Some have also claimed that the effects are more pronounced at $z>2$ \citep{digby_2010,lehmer_2013,alberts_2016}. 

Nonetheless, the bulk of the literature mentioned above involves extensive redshift ranges and sizable samples displaying diverse cluster morphologies. This diversity inherently complicates the identification of the specific physical processes that could influence the observed behaviors. Both AGN fueling and star-formation largely depend on the availability of the gas supply to the galaxy and require some transport mechanism to remove the fuel's angular momentum. Ram pressure stripping \citep{gunn_1972,ebeling_2014,poggianti_2017}, evaporation by the hot interstellar medium \citep{cowie_1977}, and tidal harassment \citep{farouki_1981,moore_1996} are all known to exhaust the cold gas reservoirs in galaxies.~Yet, some studies proposed that ram pressure from the ICM can forcibly drive the gas towards the center of the galaxies, thereby triggering AGN activity among the cluster members \citep{ebeling_2014,poggianti_2017_b}.

Cluster mergers can perturb the dynamically relaxed cores of the ICM and can lead to changes in the density, temperature, and metallicity distribution \citep{kapferer_2006}. Several studies at low to moderate redshifts have proposed that merger-driven shock waves in massive clusters trigger significant star-formation and AGN activity, with subsequent quenching in a few hundred Myr \citep{sobral_2015,stroe_2015,stroe_2017,stroe_2021}. Some studies also listed galaxy-galaxy interactions as a key driving mechanism for AGN triggering in ``disturbed" clusters \citep[e.g.,][]{owen_1999,miller_2003,noordeh_2020}. However, it is less clear whether this phenomenon persists at high redshift, where cluster formation is in a more active phase \citep{poole_2007}.

In this study, we utilize \chandra\ X-ray observations to investigate nine high-redshift galaxy clusters from the Massive and Distant Clusters of \textit{WISE} Survey (\madcows, \citealt{Gettings_2012}; \citealt{stanford_2014}; \citealt{brodwin_2015}; \citealt{mo_2018}, \citealt{gonzalez_2019}, \citealt{moravec_2020}, \citealt{dicker_2020}, \citealt{decker_2022}).~\madcows\ aims to identify the most massive, infrared-selected galaxy clusters at $0.76\leqslant z \leqslant 1.5$ over the full extragalactic sky based on data from the \textit{Wide-field Infrared Survey Explorer} (\textit{WISE}) mission \citep{wright_2010}.~We morphologically classify these clusters based on their X-ray emission and investigate their AGN populations, aiming to understand the influence of the large-scale environment on AGN activity.~In \S \ref{sec:data}, we present our cluster sample and discuss the AGN selection technique. \S \ref{sec:morph} defines the various cluster morphology parameters and our choice of asymmetry statistic. We present our results in \S \ref{sec:results} and discuss the implications in \S \ref{sec:discussion}.~Throughout this paper, we assume a flat cosmology with $\Omega_{\textup{m}} =0.307$, $\Omega_{\Lambda} =0.693$, and H$_{0} = 67.7$ km s$^{-1}$ Mpc$^{-1}$ estimated from the Planck Collaboration \citep{planck_2016} to calculate the distances and the luminosities. We define $r_{500}$ as the radius where the enclosed average mass density is equal to 500 times the critical density of the universe at the redshift of the cluster, and M$_{500}$ is the corresponding enclosed mass.

\section{Sample and Data Reduction} \label{sec:data}
\subsection{Cluster Sample}
Our pilot study includes nine massive clusters of galaxies from \madcows\ with redshifts $0.819 \leqslant z \leqslant 1.230$ and M$_{500}$ ranging from $1.8\times 10^{14}$ \(\textup{M}_\odot\) to $6\times 10^{14}$ \(\textup{M}_\odot\).~These are some of the most massive clusters known at this epoch and are intriguing in the sense of their dynamic activity.~Several of these clusters look like mergers in their X-ray morphology \citep[e.g., MOO J1142+1527 from][]{ruppin_2020}.~To date, there are ten \madcows\ clusters observed with \chandra; however, we exclude MOO J1155+3901 from our sample because the \chandra\ data do not have sufficient signal-to-noise ratio (S/N) to robustly conduct our analysis.~All were observed with the Advanced CCD Imaging Spectrometer (ACIS) chips on board \chandra\ in the VFAINT data mode. We obtained the bulk of the ACIS-I imaging observations from the publicly available \chandra\ Data Archive (PIs: Brodwin, Stanford) except for MOO J1046+2758 and MOO J1059+5454, which were imaged during Cycle 22 (PI: Ruppin). MOO J1229+6521 is also identified as PSZ2 G126.57+51.61 in the Planck survey and has multiple ACIS-S observations from Cycle 21 (PI: Bartalucci).~Therefore, we only use these newly archived ACIS-S images instead of the earlier $\sim9$ ks single ACIS-I observation for a better resolution. We list the general information of the clusters and the \chandra\ data in Table \ref{tab:data}\footnote{A complete list of \chandra\ observations used in this study can be found here:\dataset[doi:10.25574/cdc.232]{https://doi.org/10.25574/cdc.232}.}.

\input{table1}

\subsection{Data Reduction}\label{subsec:data_reduc}
Our cluster images were produced by the \chandra\ pipeline processing from level--1 event files, which we reduced using {\tt\string CIAO} \footnote{See \url{https://cxc.harvard.edu/ciao/} for details on {\tt\string CIAO.}} version 4.17 with the appropriate gain maps and calibration products based on the calibration database ({\tt\string CALDB}; version 4.11.6), provided by the \chandra\ X-Ray Center (CXC). We have reprocessed the level--1 event files using the {\tt\string chandra\_repro} script, which corrects charge transfer inefficiency and creates a new bad pixel file and a new flare-cleaned level--2 event file for each observation.~To create exposure-corrected images, exposure maps, and point-spread function (PSF) maps from the level--2 event files for single observations, we have used the {\tt\string fluximage} script only including the active ACIS chips (chips I0–I3).~We imposed a 0.7--7.0 keV bandpass that optimizes the ratio of the cluster to background flux for our purposes.~For multiple observations, we have used the {\tt\string merge\_obs} script that first creates individual images, exposure maps, and PSF maps for each observation, then reprojects into a common aspect solution, and finally combines to single images/exposure maps/PSF maps.~We generated exposure and PSF maps using an effective energy of 3.0 keV. For each exposure map, a corresponding effective exposure time map was generated by normalizing the map to its maximum effective area and scaling it by the exposure time. The PSF size was determined based on the 80\% enclosed counts fraction for each pixel location in the PSF map images.~Data products used to generate catalogs, figures, and relevant results in this study are additionally accessible from Zenodo:~\dataset[doi:10.5281/zenodo.14928038]{https://doi.org/10.5281/zenodo.14928038}.

\subsection{Cluster X-ray Masses}
Cluster masses interior to r$_{500}$ were measured following \cite{andersson_2011}, based on the Y$_X$--M scaling relation from \cite{vikhlinin_2009}.~Assuming an initial value of $r_{500} = 1$\,Mpc, we measure Y$_X$ ($\equiv$ M$_g \times$ kT) within this aperture, using a core-excised region for the temperature (0.15--1.0$r_{500}$). From the measured value of Y$_X$ and the Y$_{X,500}$--M$_{500}$ scaling relation, we can estimate a new value of M$_{500}$ and r$_{500}$ ($\equiv$ (M$_{500}/(500\times \frac{4}{3}\pi\rho_{crit}))^{1/3}$). We measure Y$_X$ within this new radius and repeat the process until r$_{500}$ converges to a stable value, yielding an estimate of M$_{500}$.

\subsection{Point Source Detection} \label{sec:detect}
To identify X-ray point sources, we initially ran {\tt\string wavdetect} \citep{freeman_2002} on the 0.7--7.0 keV energy threshold images following \cite{Vikhlinin_1998}. We used wavelet scales of \texttt{"1 2 4 8 16"} pixels, which is a reasonable default for \chandra\ data completeness as prescribed by CXC. Since our images suffer from low photons/pixel ratios, we have adopted a false-positive probability threshold of $10^{-5}$, which is slightly liberal for detecting fainter sources. However, this allows {\tt\string wavdetect} to include a fair number of spurious sources, particularly associated with cluster ICM, that require special treatment to eliminate.  

\subsection{Sensitivity Maps}
\label{subsec:sensitivity}
For a given \chandra\ observation (single/merged), the minimum flux required to detect a point source reliably depends on the location of the source in the detector plane, the size and shape of the local PSF, the local background, the integrated exposure time, and the vignetting corrections.~We followed \cite{ehlert_2013} to determine flux limits at each pixel location by calculating the minimum number of counts required to satisfy a specified threshold.~This threshold is typically defined by the binomial no-source probability ($P_B$), which quantifies the likelihood of observing the same or a higher number of counts at the source location purely due to background fluctuations.~For $S$ counts in the source extraction region $\Omega_{src}$, and $B_{ext}$ counts in the external background extraction region $\Omega_{ext}$, $P_B$ can be obtained by:

\begin{equation} \label{eq:0}
P_B (X \geqslant S) = \sum_{X=S}^{N} \frac{N!}{X!(N-X)!} p^{X} (1-p)^{N-X}
\end{equation}

\noindent where, $N=S+B$, and $p=1/(1+\Omega_{ext}/\Omega_{src})$. 

To estimate the background level at each pixel, we created background maps from the 0.7--7.0 keV cluster images by masking all point sources detected using {\tt\string wavdetect} and refilling the masked regions with random counts sampled from the local surroundings.~These background maps were then used to calculate the minimum source counts required for detection by solving Equation \ref{eq:0} based on a given $P_{B}$ criterion.~To ensure both completeness and reliability of our sample, we empirically set $P_B<0.004$ and accounted for local PSF variations using the corresponding PSF maps for each image.~The derived minimum counts were converted into count rates using the effective exposure time maps (\S\ref{subsec:data_reduc}), which were subsequently converted into limiting physical fluxes under the assumption of an absorbed power-law model with photon index $\Gamma=1.4$ and galactic hydrogen column densities ($N_{\textup{H}}$) listed in Table \ref{tab:data}.~We have used the {\tt\string Colden} \footnote{\url{https://cxc.harvard.edu/toolkit/colden.jsp}.} tool from CXC's Proposal Planning Toolkit to estimate $N_{\textup{H}}$ in each sky direction.~The choice of $\Gamma$ is optimized to be consistent with the \chandra\ Deep Field-South (CDF-S) studies \citep{xue_2011,lehmer_2012,luo_2017}. 

\begin{figure*}[htbp!]
\includegraphics[scale=0.3]{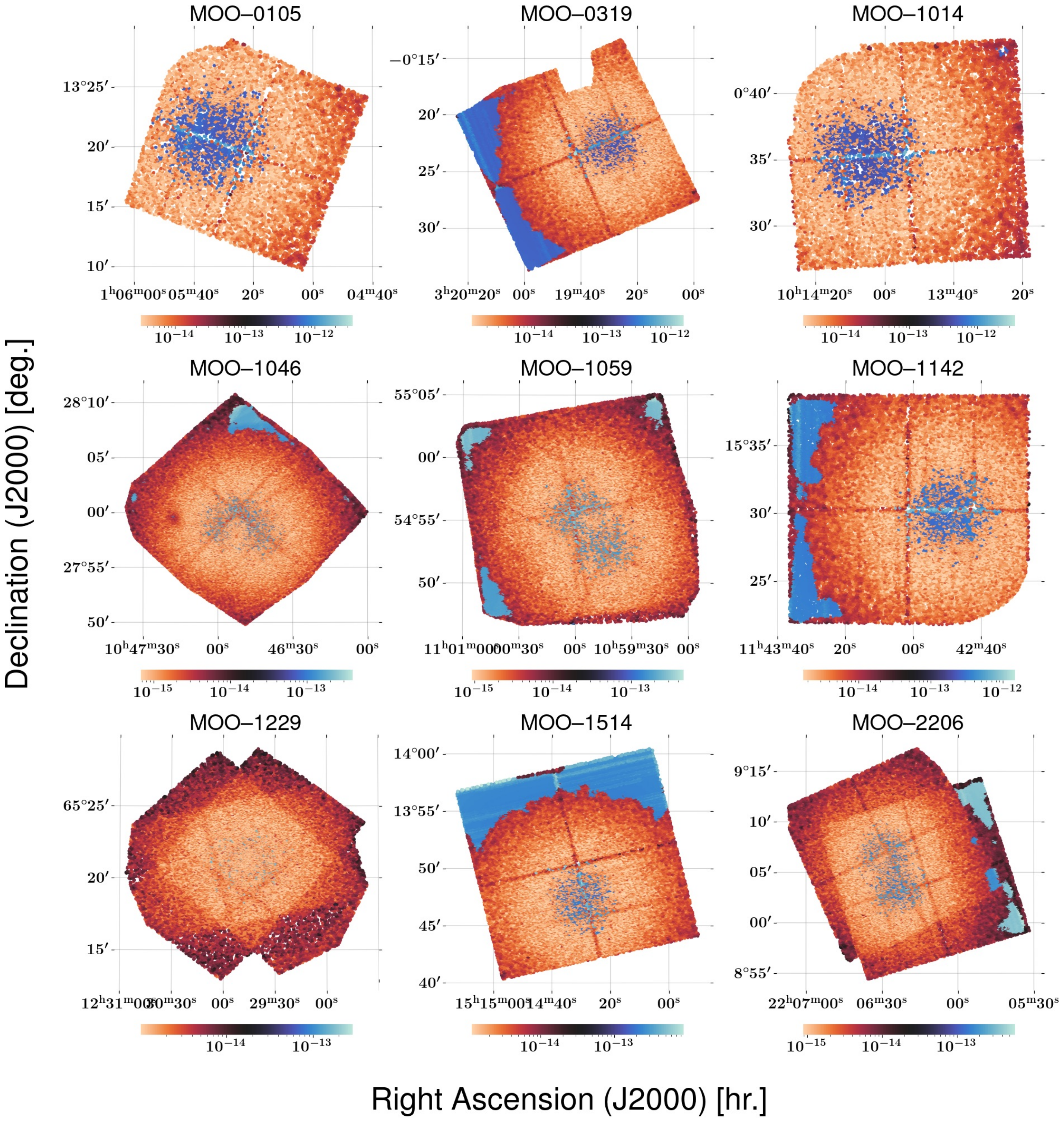}
\centering
\caption{Sensitivity maps for the nine \madcows\ cluster fields in the 0.7-7.0 keV energy band.~These maps represent the physical flux values (in units of erg cm$^{-2}$ s$^{-1}$) required for source detection, calculated by determining the number of counts needed to achieve a binomial probability of $P_{b}=0.004$. For these calculations, we assumed an absorbed power-law spectrum with a photon index of $\Gamma=1.4$. \label{fig:sensitivity}}
\end{figure*}

Figure \ref{fig:sensitivity} presents the flux sensitivity maps for each cluster field, where unreliable pixels (particularly near the detector edges) were intentionally excluded to maintain purity.~The variation in flux across the detector plane is evident, emphasizing that the flux limits are substantially higher in regions with diffuse cluster emission compared to the surrounding areas.~For each cluster field, we determined a cluster-specific flux limit as the minimum flux to which 50\% of that field is sensitive; these values are listed in Table \ref{tab:completeness}.

\input{table2}

\subsection{Final AGN Selection and Control Fields}
\label{subsec:final_catalog}
To remove spurious sources, we refined the initial source lists generated by {\tt\string wavdetect} using sensitivity maps.~Source fluxes were calculated by first estimating counts ($S$) within the source PSF regions from 0.7–7.0 keV images using {\tt\string dmextract}.~The counts were then converted into energy fluxes following the procedure described in Section~\ref{subsec:sensitivity}.~A source detection was considered reliable if its flux exceeded the local limiting flux threshold defined by the sensitivity map.~To further enhance the purity of the final selection, we applied additional cluster-specific flux limits from Table \ref{tab:completeness}.~All selected sources were visually inspected to exclude instances of diffuse cluster emission, spurious detections, and artifacts near the detector edges, which were removed if identified.~Ultimately, 518 out of the 811 sources detected by {\tt\string wavdetect} (~64\%) across all cluster fields met these criteria.~We identify these sources as AGN because their faintest luminosity in the 0.7–7.0 keV band, $7.6\times10^{42}$ erg/s, exceeds the typical 0.5–8 keV luminosity threshold of $3\times10^{42}$ erg/s for star-forming galaxies \citep{bauer_2004A}.~Accordingly, we adopt $L_{X} = 7.6\times10^{42}$ erg/s as the luminosity cutoff for our survey to maximize the retention of AGNs.

In this study, we consider AGNs located within $3r_{500}$ of the cluster centers as being associated with the cluster (or as potential interlopers), while those located at $>3r_{500}$ are used as our primary control samples (except for MOO J1229+6521, where the limited field of view of the ACIS-S chip does not extend beyond $3r_{500}$).~As a secondary control field, we use the CDF-S AGN number counts, based on comparable 0.5–8.0 keV data \citep[][hereafter L12]{lehmer_2012}.~We use this additional control field as a check for reliability.

\subsection{Completeness Correction}
To assess the detection completeness at the level of the survey luminosity cut in both the central $3r_{500}$ regions and the $>3r_{500}$ regions for each cluster, we conducted a series of mock ACIS observations with properties akin to the real observations using the {\tt\string MARX} simulator, following the approach outlined by \cite{xue_2011}.~These mock observations are derived from real observations by masking and replacing point sources with local background events.~Subsequently, we introduce a predetermined number of mock point sources at random flux levels, including those as low as our designated luminosity cutoff.~We then apply the same detection criteria outlined in \S \ref{sec:detect} to determine the number of these introduced sources detected within both the $\leqslant 3r_{500}$ and $>3r_{500}$ regions in the 0.7--7.0 keV band. By repeating the simulations at least 100 times, with 10 iterations for each run (totaling $\sim1000$ iterations), we can ascertain the completeness levels from the fraction of the input sources detected in a given region above the survey luminosity cut, as summarized in Table \ref{tab:completeness}.~Note that, due to the small field of view of MOO J1229+6521, we have used the completeness level from the CDF-S survey ($\sim90\%$ at our survey luminosity cut) as a proxy for its local field completeness.~Unless stated otherwise, we will utilize these values in our subsequent analysis to correct for incompleteness throughout the article.

\section{Cluster Morphology} \label{sec:morph}
\subsection{Classical Morphology Parameters}
X-ray images can generally detect two distinctive features of the ICM that indicate a non-virialized cluster: (1) the presence of elongated substructures, and (2) X-ray centroid variation. Dynamically relaxed clusters usually exhibit cool cores with central sharp X-ray peaks. Conversely, a departure from axial symmetry indicates the presence of secondary peaks or substructures.

To quantify the cluster dynamical state, we need a robust indicator.~Several estimators have been proposed in the past; power ratios \citep{buote_1995,buote_1996,jeltema_2005} and emission centroid shifts \citep{mohr_1993,poole_2006,bohringer_2010} are among the most common ones.~Other studies use asymmetry, smoothness and concentration \citep{raisa_2013}, Gini, and $M_{20}$ \citep{parekh_2015}\footnote{For a more comprehensive review of various substructure statistics and their applications, refer to \cite{weibmann_2013b}, \cite{chong_2016}, \cite{lovisari_2017}, and \cite{green_2019}.}.~However, most of these are designed for use in only a narrow range of applications, particularly at low redshifts with high S/N observations.

\subsection{Photon Asymmetry} \label{subsect:phot_asym}
We have tested several morphology parameters mentioned above to classify our cluster sample; however, none proved optimal for our case.~Most of these parameters require at least $\sim10^{5}$ counts per image, where our observations typically have only $\sim 10^{3}$--$10^{4}$ counts. In this study, we choose photon asymmetry ($A_{\text{phot}}$), a substructure statistic developed by \citet[hereafter N13]{nurgalive_2013}. It quantifies how much the X-ray emission deviates from the idealized axisymmetric case. We picked $A_{\text{phot}}$ over other estimators for the following reasons:\\
 \begin{enumerate}
    \item It differs from the standard optical asymmetry, which only works in the counts/pixel (binned) $\gg1$ regime, whereas our X-ray images have $\sim10^{-2}$ counts/pixel. In principle, we could use smoothed images to alleviate the situation; however, that would have introduced systematic offsets (see the discussion in \S 4.1 of \citetalias{nurgalive_2013}).
    \item It is suitable for high redshifts and a wide range of S/N ratios \citep[hereafter N17]{nurgaliev_2017}.
    \item It is independent of exposure times and background levels.
    \item It is observationally well-motivated and the only estimator that remains unbiased.\\
 \end{enumerate}
 
For the full description of $A_{\text{phot}}$, we refer the reader to \citetalias{nurgalive_2013}.~In short, it splits the image into a few user-defined annuli and checks whether the surface brightness is uniform in each of those. To assess the degree of non-uniformity of the angular distribution of the photons, it uses the non-parametric Watson’s test \citep{watson_1961} to estimate the distance between the true cumulative photon distribution $(F)$ and a uniform cumulative distribution function $(G)$ that represents an idealized axisymmetric source.~Thus, for a single annulus with $N$ total counts and $C$ cluster counts (estimated by subtracting the expected number of background counts in that annulus from $N$), the distance$(F, G)$:

\begin{equation} \label{eq:1}
\hat{d}_{N,C} = \frac{N}{C^{2}} \bigg(U_{N}^{2}-\frac{1}{12}\bigg)
\end{equation}
Here, $U_{N}^{2}$ is the Watson's statistic \citep{watson_1961}.~In theory, the greater values of $\hat{d}_{N,C}$ signify more dynamically disturbed clusters. For a $k$ number of annuli, photon asymmetry calculates the weighted sum of distances $\hat{d}_{N_{k},C_{k}}$ from each annulus with a weight equal to the estimated number of cluster counts $C_{k}$ in that annulus:

\begin{equation} \label{eq:2}
A_{\text{phot}} = 100 \sum_{k=1}^{k} C_{k} \hat{d}_{N_{k},C_{k}}\bigg/\sum_{k=1}^{k} C_{k}
\end{equation}
The multiplication factor 100 in the above equation ensures that photon asymmetry can have values in the range $0 < A_{\text{phot}}  \lesssim 3$, where higher values indicate more asymmetry. 

\begin{figure*}[bthp]
\includegraphics[scale=0.65]{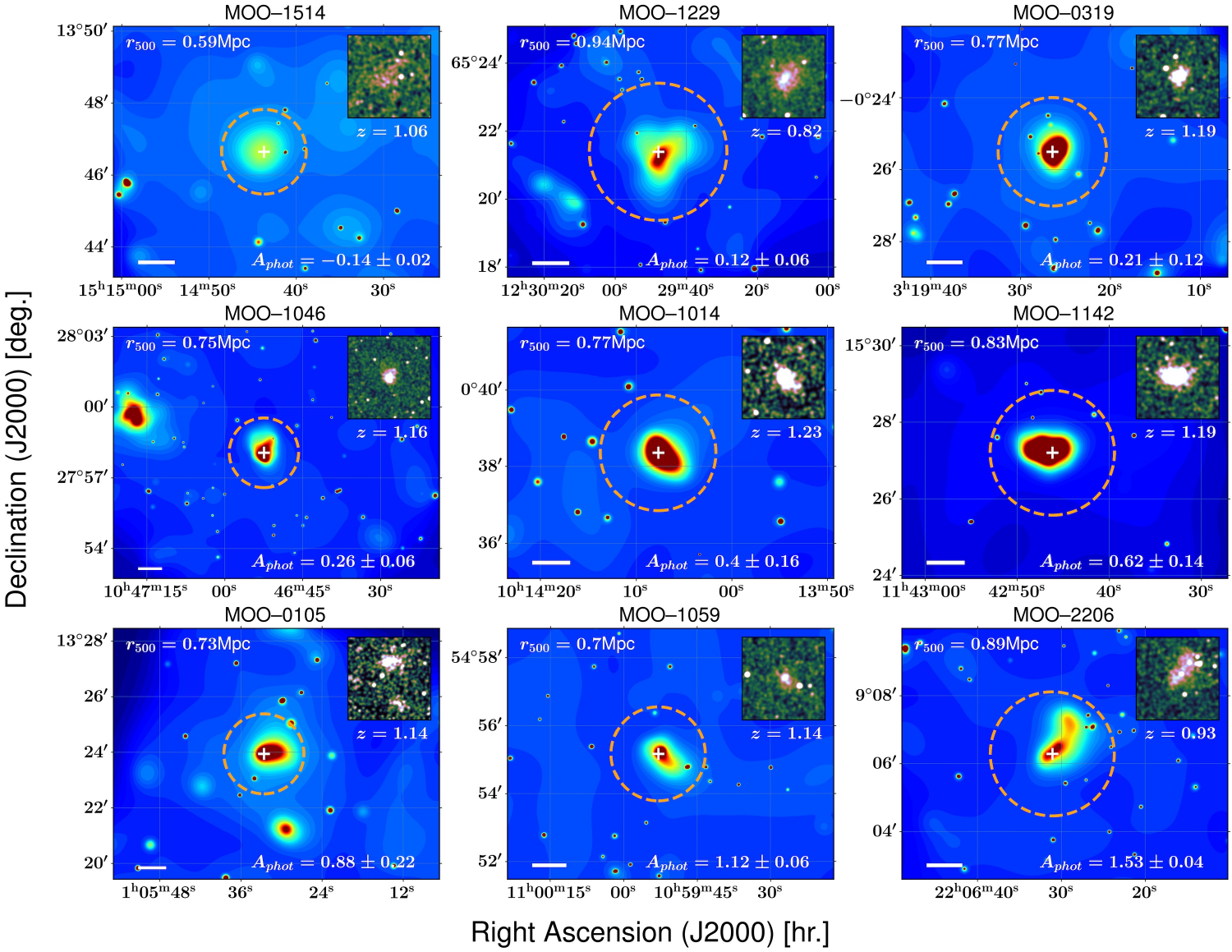}
\centering
\caption{Adaptively-smoothed 0.492 arcsec/pixel binned X-ray images of the clusters, ranked by $A_{\text{phot}}$.~As discussed in \S \ref{subsect:phot_asym}, We applied the smoothing conditions only to suppress the noise and highlight the presence of substructure, they were not used to calculate $A_{\text{phot}}$. In the insets, we show 0.984 arcsec/pixel binned exposure-corrected 0.7--7.0 keV images that has been smoothed with a fixed-width Gaussian filter with $\sigma=4 \arcsec$. The dashed orange circles outline $r_{500}$ circles, which are centered on the brightest X-ray pixels after point source marking (white ``+” signs). The size of the scale bar in the bottom left corner in each image is 0.5 Mpc. The figure demonstrates that as the values of the substructure statistics increase from top-to-bottom and left-to-right, the clusters appear more and more dynamically disturbed. Errors in $A_{\text{phot}}$ are $1\sigma$ and estimated from bootstrap subsampling method following \citetalias{nurgalive_2013}. \label{fig:aphot}}
\end{figure*}

To calculate $A_{\text{phot}}$, we have used flare-cleaned 0.7--7.0 keV threshold images of our cluster sample. We replaced each point source with interpolated values obtained from their local background regions using the {\tt\string CIAO} routine {\tt\string dmfilth}.~We do this to ensure that the X-ray peaks are from the over-dense regions within the cluster ICMs rather than AGNs. We have chosen the brightest X-ray pixels as cluster centers and $r_{500}$ circles as maximum apertures with radial binning of 0.05, 0.12, 0.2, 0.30, and 1.0 $r_{500}$, as prescribed by \citetalias{nurgalive_2013}. We estimated sky backgrounds using annuli outside of the $3r_{500}$ radii in each image to avoid any contamination from the cluster emission.~Our cluster ranking based on increasing $A_{\text{phot}}$ is presented in Figure \ref{fig:aphot}. We use the {\tt\string CIAO} routine {\tt\string csmooth} to adaptively smooth these images only to suppress noise and highlight the presence of substructure; we do not use the smoothed images to calculate $A_{\text{phot}}$. To provide perspective of the physical scale, the dashed orange circles outline the $r_{500}$ radii centered on the cluster X-ray centroids in each. We estimated the $1\sigma$ errors from the bootstrap subsampling method following \citetalias{nurgalive_2013}. 

As one can see, the clusters appear more and more asymmetric from the top-to-bottom and left-to-right in Figure \ref{fig:aphot}, and the increasing $A_{\text{phot}}$ values mostly match the ``by-eye" ranking.~\citetalias{nurgaliev_2017} noted $A_{\text{phot}}<0.15$ and $A_{\text{phot}}>0.6$ as reasonable (not strict) thresholds for relaxed and disturbed cluster morphology, which is in agreement with our observations.~Indeed, \cite{ruppin_2020} identified MOO J1142+1527 as an ongoing merger that hosts a cool core at the location of the X-ray peak.~Similarly, from our images, it is apparent that MOO J0105+1323 and MOO J2206+0906 are early-stage mergers since both manifest secondary X-ray peaks.~This is less clear for MOO J1059+5454; although, the asymmetric structure along with higher $A_{\text{phot}}$ suggest it might be a major merger within the past $\sim$1–2 Gyr (\citetalias{nurgaliev_2017}).~Another interesting case is MOO J1046+2758, which appears to have an apparent neighbor in the northeast direction. We identify this cluster as RM J104716.9+275926.1 at a redshift of 0.43 \citep{Rykoff_2016}, and therefore consider it physically unassociated. We will assume clusters with $A_{\text{phot}}>0.6$ possess disturbed morphology for this study. 

\citetalias{nurgalive_2013} quotes one small caveat: when the square root of the total counts becomes comparable to the cluster counts, the estimated cluster counts can become $\sim$zero, or even negative. As a consequence, $A_{\text{phot}}$ can have values outside of the defined range, which we suspect is the primary reason for observing the negative value of $A_{\text{phot}}$ in MOO J1514+1346 since its image contains a substantial amount of sky noise.~Another pitfall is that the choice of radial binning depends slightly on the cluster itself; thus, optimizing a single binning scheme for the entire sample for consistency can be problematic. As a result, a low number of $C$ in any particular bin can lead to additional systematic uncertainties, which may have been overlooked by the quoted $1\sigma$ errors estimated from bootstrap subsampling.

\section{Results}\label{sec:results}
\subsection{Dynamical State of the ICM}\label{subsec:dynamical_state}
\begin{figure*}[htbp]
\includegraphics[scale=0.3]{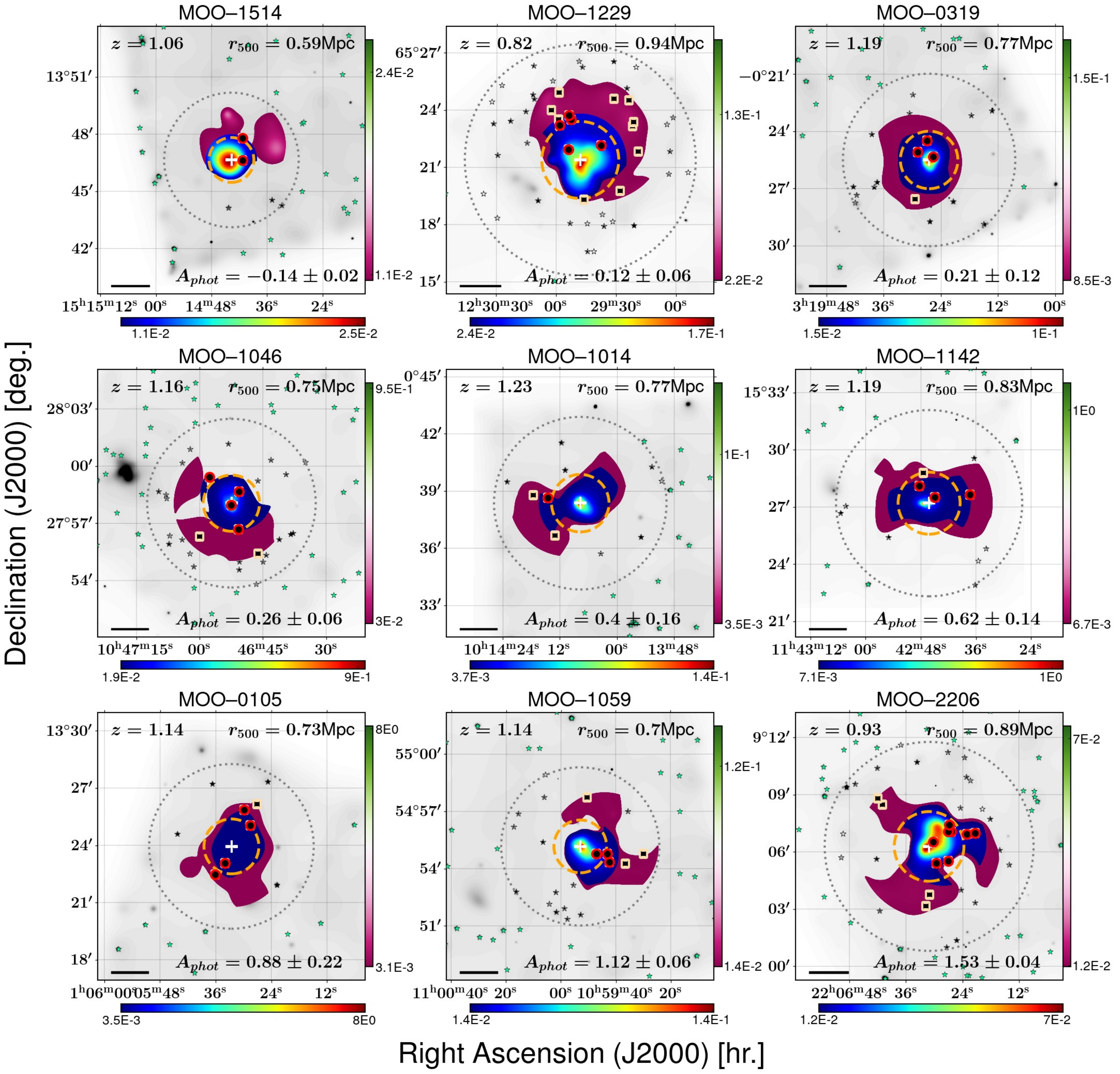}
\centering
\caption{In grayscale are the adaptively-smoothed X-ray images of the clusters ranked by $A_{\text{phot}}$, similar to Figure \ref{fig:aphot}.~In addition, to emphasize ICM structures, The ``jet" colormaps (bluish) cover equivalent isophote areas bound by the $r_{500}$ circles (dashed orange) in each cluster, while the ``purple" colormaps stretch up to 1.5 times the area.~AGNs that meet the selection criteria outlined in \S\ref{subsec:final_catalog} are marked with open stars for regions within $3r_{500}$ (dotted gray circle) and green-filled stars for regions beyond $3r_{500}$.~The red circles and beige squares denote AGNs detected within $1.0r_{500}$ and  $1.5r_{500}$ isophote areas, respectively. The color bars indicate counts per pixel, with white ``+" signs marking the locations of X-ray peaks. The scale bars represent a physical size of 1.0 Mpc.\label{fig:aphot_masked}}
\end{figure*}

We have demonstrated in Figure \ref{fig:aphot} that the photon asymmetry visually matches our expectation of quantifying the dynamical state of the ICM. We explore this further in Figure \ref{fig:aphot_masked}. Apart from ranking clusters based on $A_{\text{phot}}$, we trace their ICM structures by examining their isophote contours in adaptively smoothed point source masked images. The ``jet" colormaps (bluish) highlight isophote surfaces with equivalent areas bound by the $r_{500}$ circles (orange) that are centered on the X-ray peaks (white ``+") of each cluster, while the  ``purple" maps underneath stretch up to 1.5 times the equivalent area.~The advantage of this approach is that instead of including low surface brightness regions that may not be part of the clusters, we focus more on following the distributions of the respective ICMs while using an area equivalent to that enclosed within $r_{500}$.~We mark the AGNs within $1.0r_{500}$ and $1.5r_{500}$ isophote areas as red circles and beige squares, respectively. Figure \ref{fig:aphot_masked} underlines two important results:  
 \begin{enumerate}
    \item AGNs seem to preferentially lie in the over-dense cluster regions, especially within isophotes, although we expect at least a few of these to be simply in the cluster line-of-sight since we do not know their membership status. 
    \item There is possible evidence that some clusters host a higher number of AGNs than others (for example, compare MOO J2206+0906 to MOO J1514+1346).
    
 \end{enumerate}

\subsection{AGN Incidence vs. ICM Surface Brightness}
\begin{figure*}[htbp]
\includegraphics[scale=0.33]{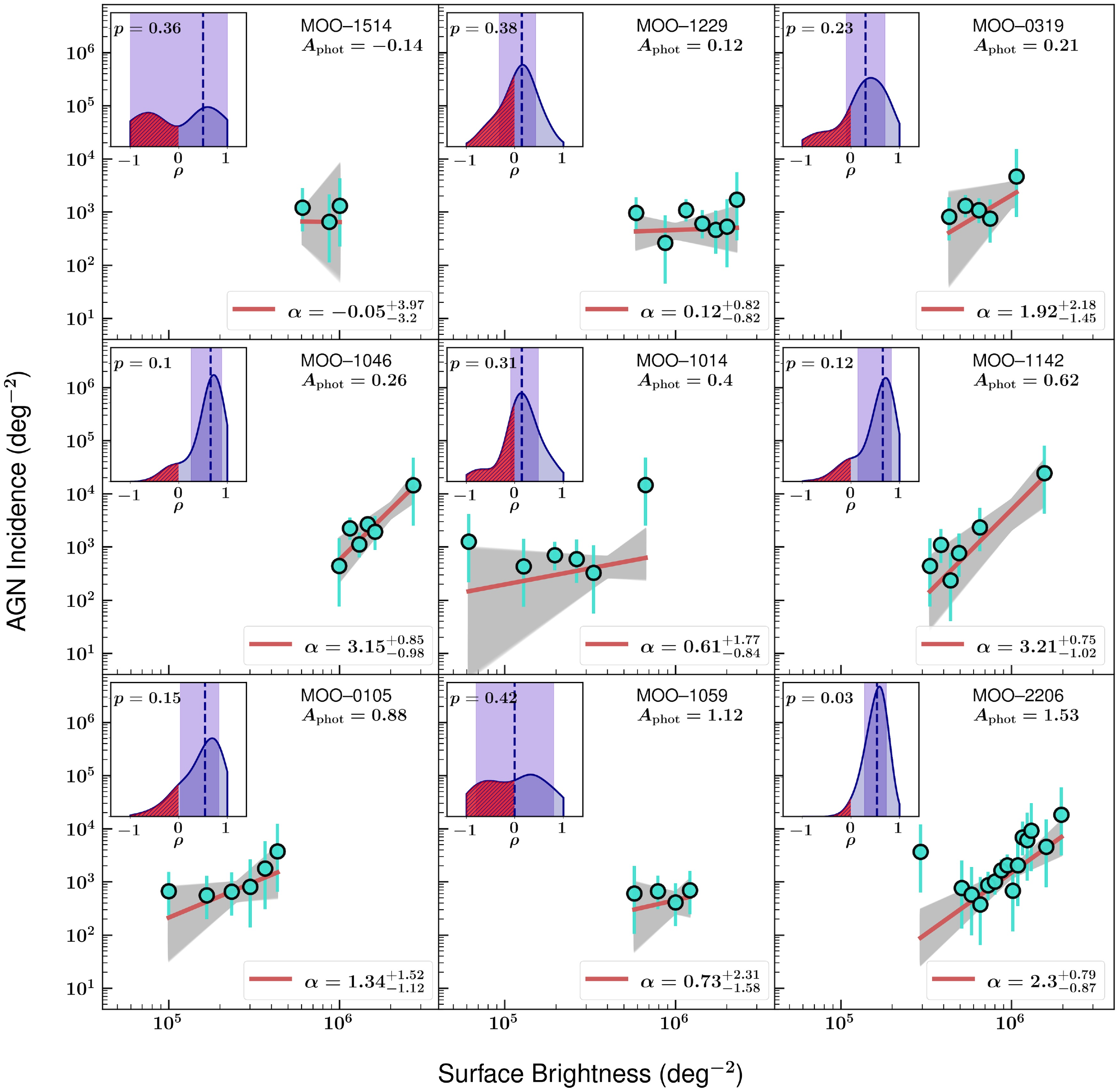}
\centering
\caption{Double-logarithmic plots of X-ray surface brightness within $\leqslant3r_{500}$ clusters versus completeness-corrected projected AGN incidence, including $1\sigma$ Poisson errorbars. The brown line represents the best-fit power-law models derived from MCMC with corresponding $1\sigma$ confidence regions in light gray. The insets depict the KDEs of the Spearman's rank correlation coefficients from Monte Carlo simulations with medians (50th percentiles) marked by the vertical blue-dashed lines and the $1\sigma$ confidence regions (i.e., between 16th and 84th percentiles) in light-violet. The red-hatched regions highlight the p-value areas for rejecting the null hypothesis.\label{fig:agn_sb}}
\end{figure*}

To further emphasize the points discussed above, we slice the entire $3r_{500}$ circular areas into small square units $(\sim17\arcsec$--$29\arcsec)$ and compute the local 0.7--7.0 keV X-ray surface brightnesses (counts/deg$^{2}$) as well as projected AGN incidence (AGNs/deg$^{2}$).~Using this approach, we explore the relationship between cluster surface brightness and completeness-corrected AGN incidence (at our survey luminosity cut, as listed in column 3 of Table \ref{tab:completeness}) in Figure \ref{fig:agn_sb}.~Assuming a simple power-law distribution, we show the best-fit model (in brown) estimated from the Affine Invariant Markov chain Monte Carlo (MCMC) package {\tt\string emcee} \citep{foreman_2013} along with corresponding $1\sigma$ confidence regions in light gray.  

Figure \ref{fig:agn_sb} demonstrates that in certain clusters, AGNs may preferentially reside in regions with high ICM density, as suggested by the non-zero power-law indices ($\alpha$) derived from the MCMC analysis.~For instance, clusters such as MOO J1046+2758, MOO J1142+1527, and MOO J2206+0906 exhibit steep profiles with greater significance.~In contrast, MOO J1514+1346, MOO J1229+6521, and MOO J1059+5454 show indices that are more consistent with no observable trend.

To confirm the presence of any statistically significant correlation, we calculate the non-parametric Spearman's rank correlation coefficients ($\rho$) that do not assume a linear relation.~The value of $\rho$ ranges from $-1$ to $+1$, where  $\rho=0$ indicates no correlation, and the extremes imply an exact monotonic relationship.~To account for the measurement uncertainties, we use Monte Carlo techniques to simulate random 10,000 instances of the data to generate the Kernel Distribution Estimation (KDE) of the probability density function from the coefficients, which we show in the insets.~As one would expect, clusters such as MOO J1046+2758, MOO J1142+1527, MOO J0105+1323, and MOO J2206+0906 exhibit median correlation coefficients (indicated by the vertical dashed blue line) consistent with moderate to strong correlation ($0.54\leq\rho\leq0.66$).~However, only MOO J2206+0906 yields a p-value of 0.03 ($2.12\sigma$), suggesting a tentative potential correlation in this case.~It is important to note that this interpretation is highly dependent on the number of data points (and thus the number of AGNs) since the reliability of the p-values is strongly influenced by sample size.

At this stage, it remains unclear whether there is a definitive distinction between the two morphological classes, a question we will explore more rigorously in the following sections.~It is also possible that the observed AGN overdensities are a byproduct of the underlying galaxy overdensities in the central regions of clusters, rather than AGNs being actively triggered.~However, due to the lack of confirmed cluster memberships, we were unable to measure galaxy densities, a limitation we aim to address in a future study.

\subsection{AGN Surface Density from Model}\label{subsec:logn-logs}
\begin{figure*}[htbp]
\includegraphics[scale=0.33]{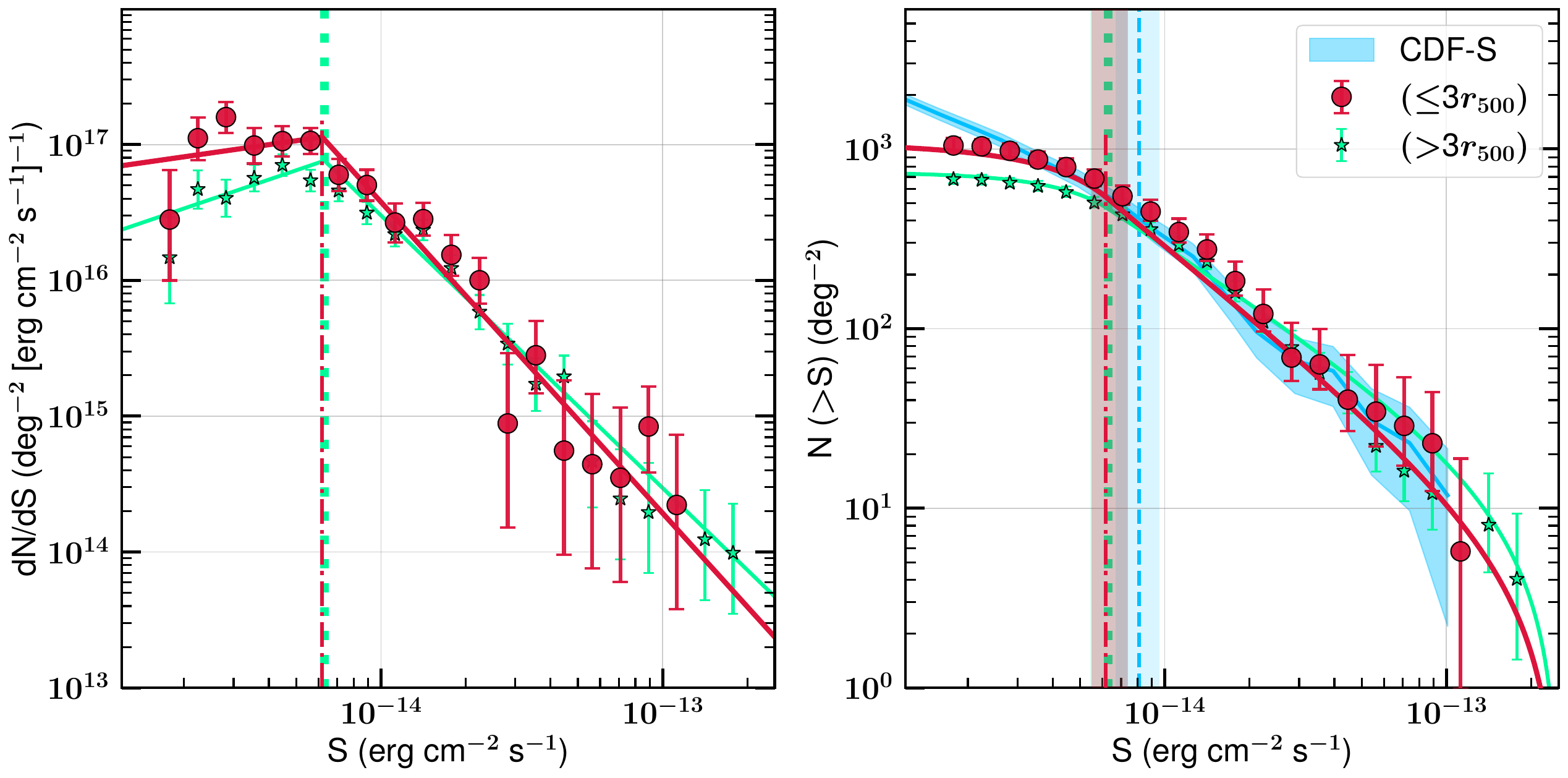}
\centering
\caption{\textbf{Left:} Differential number counts ($dN/dS$) vs.~flux ($S$, in bins of $\approx0.1$ dex) across the nine combined cluster fields in the 0.7--7.0 keV energy band. The red circles represent AGNs located within the $3r_{500}$ regions, while the light-green stars corresponds to the AGNs found outside the $3r_{500}$. The best-fit $dN/dS$ parameterizations (listed in Table \ref{tab:logN-logS}), based on Equation \ref{eq:3}, are shown as curves in their respective colors for each category. The break fluxes of the double power-law fits are indicated as vertical dashed lines in both cases.
\textbf{Right:} Cumulative number counts of AGN $(\log N-\log S)$ derived from the $dN/dS$ parameterizations. For comparison, the cumulative number counts for the CDF-S's AGNs in a comparable energy band (0.5--8.0 keV) is displayed in sky-blue \citep{lehmer_2012}. In both figures, the errorbars and color-shaded regions represent $1\sigma$ Poisson uncertainties.\label{fig:logN-logS}}
\end{figure*}

To correct for line-of-sight interlopers and to account for variability in exposure times and selection biases inherent to our flux-limited survey, here we adopt a more rigorous approach. Following the methodology outlined in Section 2 of \citetalias{lehmer_2012}, the left panel of Figure \ref{fig:logN-logS} presents the differential number counts of AGN ($dN/dS$) in flux ($S$) bins, which account for the Eddington bias near the sensitivity limit. In this plot, the red circles represent AGNs located within $3r_{500}$ regions, identified in nine cluster fields combined and selected based on the criteria described in \S \ref{subsec:final_catalog}. In contrast, the light-green stars correspond to AGNs found outside the $3r_{500}$ regions. The error bars represent $1\sigma$ Poisson uncertainties for small-number statistics, estimated following \cite{gehrels_1986}.

Past studies of X-ray number counts demonstrated that power-laws are generally good fits for the overall shapes of the $\log N- \log S$ \citep{bauer_2004A,Georgakakis_2008,luo_2017}. Therefore, we adopt the double power-law model to parametrize $dN/dS$:
\begin{equation} \label{eq:3}
    \frac{dN}{dS} = \times
    \begin{cases} 
    K(S/S_{\text{Ref}})^{-\beta_{1}} &  (S\leqslant f_{b}) \\
    K(f_{b}/S_{\text{Ref}})^{\beta_{2}-\beta_{1}} (S/S_{\text{Ref}})^{-\beta_{2}} &  (S> f_{b}) 
    \end{cases}
\end{equation}
\noindent where $f_{b}$ refers to the break flux in the double power-law model (vertical dashed lines),  $S_{\text{Ref}} \equiv 10^{-14}$ erg cm$^{-2}$ s$^{-1}$. $\beta_{1}$ and $\beta_{2}$ are the faint-end and bright-end slopes, and $K$ is the normalization. We used MCMC to derive the best-fit values for each parameter and their corresponding $1\sigma$ errors, which are provided in Table \ref{tab:logN-logS}.

\input{table3}

In the right panel of Figure \ref{fig:logN-logS}, we present the cumulative number counts of AGNs (i.e., the number of AGNs brighter than a given flux) derived from the $dN/dS$ parameterizations. For comparison, the cumulative counts of AGNs in the CDF-S within a similar energy band are also shown \citepalias{lehmer_2012}.~The vertical shaded columns indicate the $1\sigma$ error ranges for the break fluxes in each case, highlighting significant overlaps between the \madcows\ and CDF-S $f_{b}$ parameters. The fitted curves reveal that our AGNs align well with the CDF-S profile at the bright end of the flux distribution but diverge significantly at the faint end due to the negative $\beta_{1}$ parameter.~This suggests that we must rely more on completeness rather than contributions from faint sources for our survey luminosity cut $L_{X} = 7.6\times10^{42}$ erg/s.

\begin{figure*}[htbp]
\includegraphics[scale=0.66]{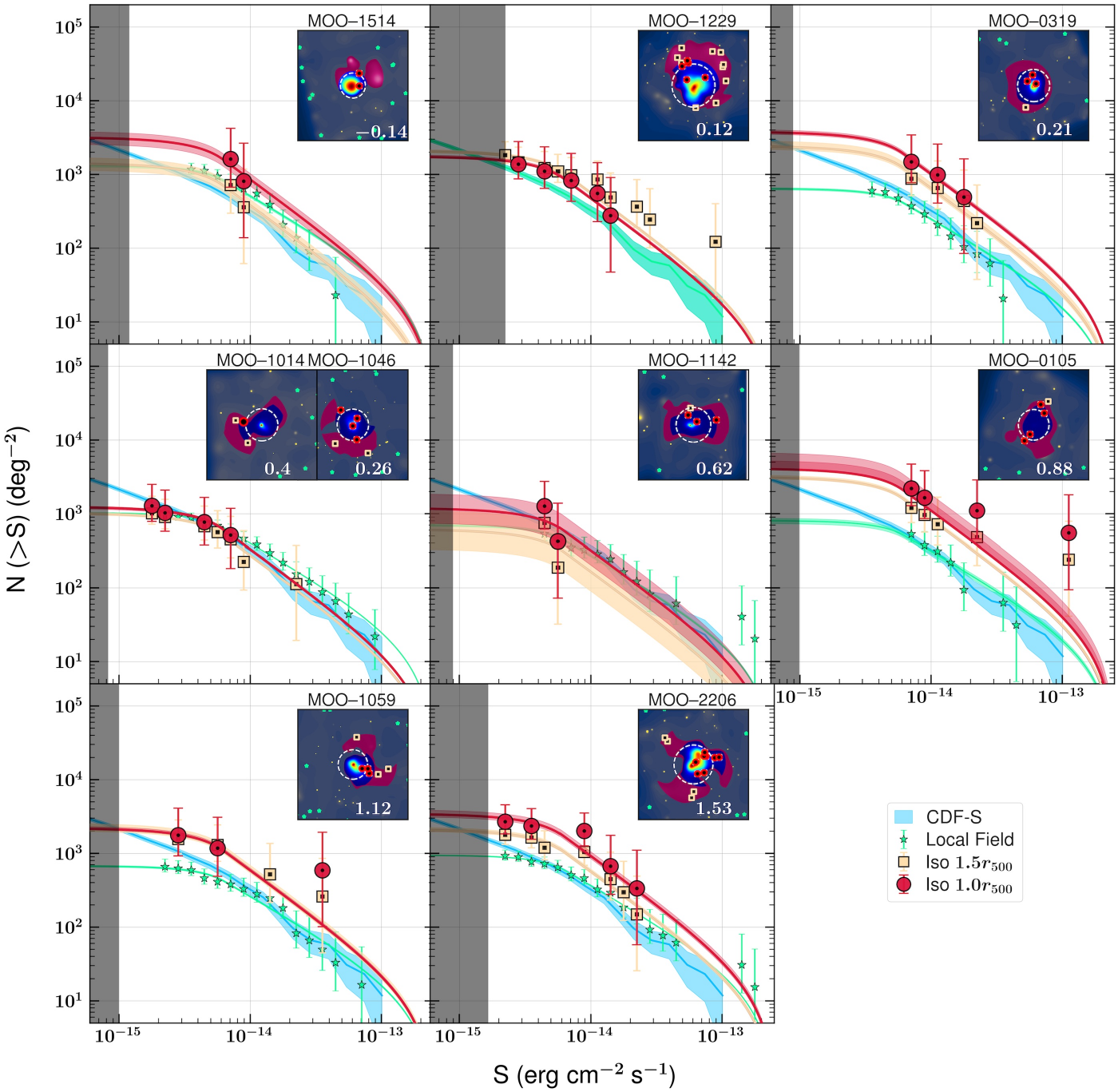}
\centering
\caption{Cumulative number counts (number of sources brighter than a given flux) with their corresponding $1\sigma$ errors displayed for the AGNs associated with the cluster in $1.0r_{500}$ (red circles) and $1.5r_{500}$ (beige squares) isophote areas, as well as for local field AGNs ($>3r_{500}$, light-green stars) in the 0.7--7.0 keV band for \madcows\ clusters. As discussed in \S \ref{subsec:logn-logs}, we combined MOO J1014+0038 with MOO J1046+2758 due to insufficient data. For each case, the best-fit $\log N- \log S$ profiles with $1\sigma$ uncertainties are represented by curves in their respective colors. The sky-blue regions correspond to the 4 Ms CDF-S cumulative number counts in the 0.5--8.0 keV energy band \citepalias{lehmer_2012}. Since MOO J1229+6521 lacks local field AGNs (\S \ref{subsec:final_catalog}), we use the CDF-S data as a proxy for the field AGN distribution, shown in light green. Insets display adaptively smoothed images overlaid with $1.0r_{500}$ and $1.5r_{500}$ isophote areas, along with their respective AGNs and quoted $A_{\text{phot}}$ values, as demonstrated in Figure \ref{fig:aphot_masked}. The dark-shaded regions on the left represent flux levels below the survey luminosity cut in each cluster field, where surface densities are not estimated from the $\log N- \log S$ profiles, as discussed in \S \ref{subsec:merger}.  \label{fig:logn-logs_iso}}
\end{figure*}

To estimate AGN surface densities at a given luminosity cut for each cluster field, we construct the $\log N- \log S$ plots separately in Figure \ref{fig:logn-logs_iso}, ordered by the $A_{\text{phot}}$ values.~In this case, however, instead of considering the entire  $3r_{500}$ radius for AGNs associated with the cluster, we restrict the analysis to $1.0r_{500}$ and $1.5r_{500}$ isophote areas (represented by red circles and beige squares, respectively; also shown in the respective inset images).~Due to insufficient data points, we combined MOO J1014+0038 with MOO J1046+2758.~This does not significantly affect the results, as both clusters have similar $A_{\text{phot}}$ values, comparable flux limits for the luminosity cut, and nearly identical $\log N- \log S$ profiles. Similarly, we plot $\log N- \log S$ for the primary control samples, i.e., the local field ($>3r_{500}$), excluding MOO J1229+6521, which are shown as light-green stars.~In the background, we also show the $\log N- \log S$ for our secondary control fields from the CDF-S in sky blue, adopted from \citetalias{lehmer_2012}.~As discussed in \S \ref{subsec:final_catalog}, the small field of view of the ACIS-S chips limits the ability to sample AGNs beyond $3r_{500}$ in MOO J1229+6521.~Therefore, we assume that the local field for this cluster is the same as that of the CDF-S. All the error bars shown are estimated from Poisson asymmetric confidence limits of \cite{gehrels_1986}.

While the X-ray luminosity functions of AGNs are known to evolve with redshift, the general shape of the $\log N- \log S$ profiles does not alter significantly within narrow redshift ranges (see \S 4.1 and Fig.~9 of \citetalias{lehmer_2012} for reference). Therefore, we assume that the shapes of the $\log N- \log S$ profiles are consistent across nine \madcows\ clusters apart from normalization, and adopt the best-fit parameters from Table \ref{tab:logN-logS}, excluding the normalization constant $K$.~We then apply a $\chi^{2}$ minimization technique to optimize the  $K$ values, thereby constructing the $\log N- \log S$ profiles for each cluster. The $1\sigma$ uncertainties in these profiles are estimated using bootstrap resampling, with the resulting confidence intervals highlighted as shaded regions in Figure \ref{fig:logn-logs_iso}.~The dark-shaded regions on the left of Figure \ref{fig:logn-logs_iso} represent flux levels below the survey luminosity cut in each cluster field, for which we do not estimate surface densities from the $\log N- \log S$ profiles.~The models illustrate that, for the survey luminosity cut, several \madcows\ clusters exhibit excess AGN relative to the primary control fields. 

\subsection{Merger Driven AGN Activity} \label{subsec:merger}
Observational studies in recent years reported that the dynamical state of the ICM may actively/passively incite AGN triggering in merging clusters \citep[e.g.,][]{moravec_2020,noordeh_2020,stroe_2021}. To fairly compare AGN incidence among clusters, we estimate AGN surface densities from the extrapolated models in Figure \ref{fig:logn-logs_iso} at our survey's luminosity cut.~This approach mitigates bias caused by variations in total exposure times across different clusters and minimizes the inclusion of very low X-ray luminosity sources unlikely to be AGNs at the cluster redshift, as well as exceptionally bright AGNs. Since cluster and field regions exhibit different completeness sensitivities for a given flux, we normalize these surface densities by their corresponding completeness fractions from Table \ref{tab:completeness}.~We define the completeness-corrected excess AGN surface density as

\begin{equation} \label{eq:4}
\Sigma_{\textup{AGN}} = \Sigma_{\textup{iso}}-\Sigma_{\textup{field}}
\end{equation}
\noindent where $\Sigma_{\textup{iso}}$ and $\Sigma_{\textup{field}}$ are the completeness-corrected AGN surface densities at the survey luminosity cut, calculated from the isophote and field regions, respectively.

\begin{figure*}[htbp]
\includegraphics[scale=0.5]{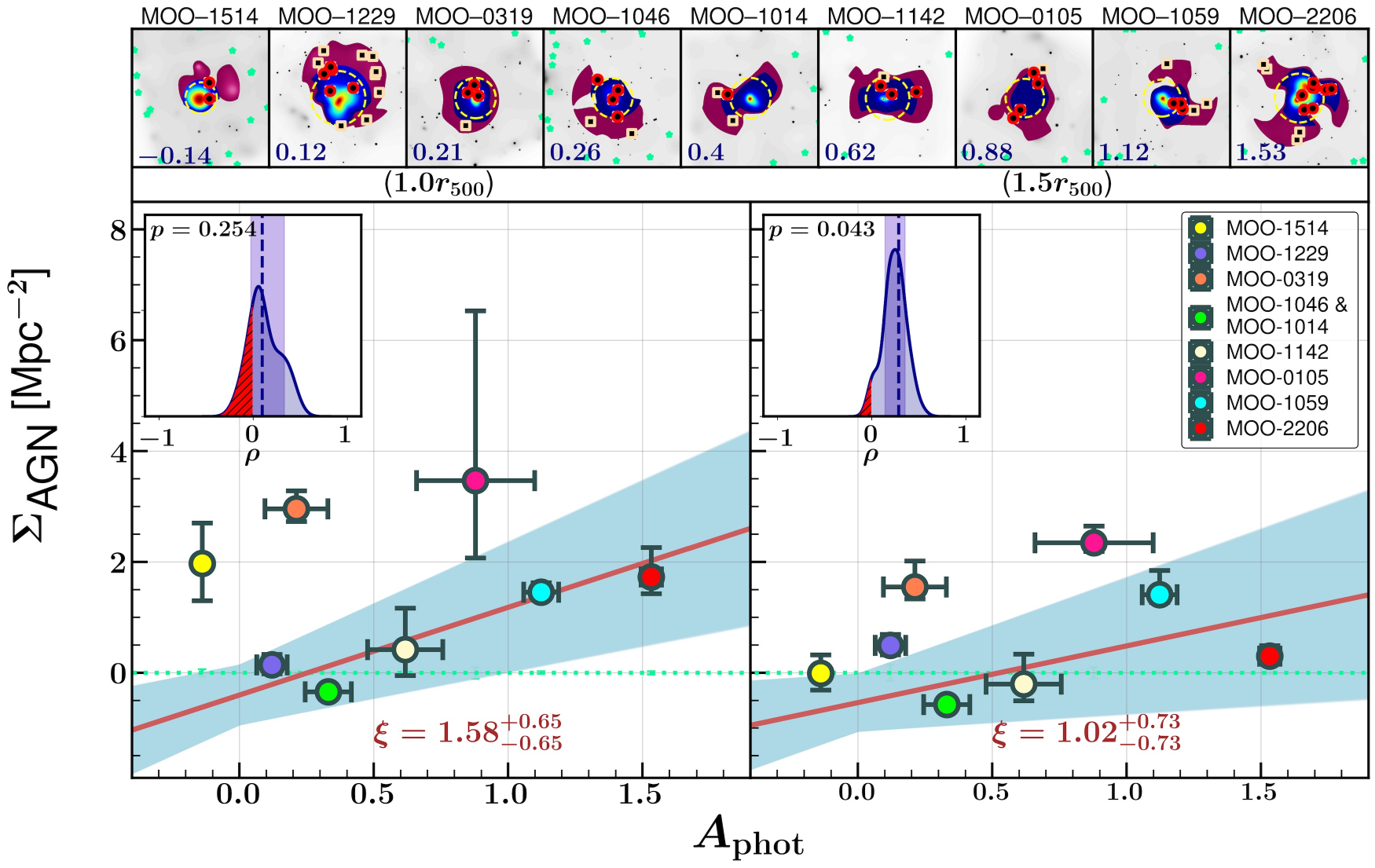}
\centering
\caption{Completeness-corrected excess AGN surface density in cluster isophotes (\textbf{left}: $1.0r_{500}$, \textbf{right}: $1.5r_{500}$) as a function of photon asymmetry, where we subtract the field expectations for each cluster estimated from the local fields ($\ge 3r_{500}$, dotted light-green lines with errorbars); hence negative AGN densities are possible.~To compute these densities, we extrapolate the $\log N- \log S$ profiles from Figure \ref{fig:logn-logs_iso} down to the survey luminosity cut and applied corrections for completeness. As before, MOO J1014+0038 and MOO J1046+2758 are combined to enhance S/N. Additionally, the best-fit linear regression model (solid brown line) with a $1\sigma$ confidence interval (light blue region) is displayed. Insets in both panels are the KDEs of Spearman’s rank correlation coefficients as before, and the upper square tiles portray the $A_{\text{phot}}$-ranked (reported in the bottom-left corner of each) cluster isophotes and their corresponding AGNs from Figure \ref{fig:aphot_masked}. All the errorbars shown represent $1\sigma$. \label{fig:surf_den_iso}}
\end{figure*}

In Figure \ref{fig:surf_den_iso}, we depict the completeness-corrected excess AGN surface densities as a function of photon asymmetry, with $\Sigma_{\textup{field}}$ derived from the local fields ($\ge 3r_{500}$).~In the left panels, we confine the isophotes to $1.0r_{500}$ equivalent areas, whereas in the right panels, we extend them to $1.5r_{500}$.~Additionally, we fit a simple linear regression model that accounts for both axes' uncertainties, which we show as a solid brown line with a $1\sigma$ confidence interval in light blue color. In both cases, the best-fit slope parameters ($\xi$) indicate non-zero slopes, albeit with large uncertainties. However, Spearman’s rank correlation suggests only a weak positive correlation (median $\rho \approx 0.3$) in the $1.5r_{500}$ isophotes, with a significance level of $\sim2\sigma$ (p-value $\approx 0.043$).~To verify that our chosen luminosity cut does not introduce significant bias into the study, we tested various luminosity cuts but found no statistically significant effect that would alter the final results.~Furthermore, these findings are consistent with cases where no completeness corrections are applied.

Despite this, when using CDF-S estimates for the field, we do not observe any correlations in either cases (not depicted in the plots).~However, this does not account for cosmic variance, which could be crucial for our study given its susceptibility to small-number statistics.~Indeed, previous studies have reported that cosmic variance emerging from large-scale density fluctuations can lead to uncertainties in observational estimates of galaxy/AGN number densities within clustered populations, potentially exceeding Poisson uncertainties \citep{somerville_2004,moster_2011}.~This effect is particularly pronounced in small fields and at high redshifts.~Consequently, a global field estimate from deep surveys like the CDF-S may not adequately represent the underlying field populations for our \madcows\ clusters, where field-to-field variation is prominent \citep{luo_2017}.~Given that our derived excess AGN surface density estimates are based on small numbers of cluster AGNs and only nine clusters in our sample, this can introduce significant bias in the observed correlation trend (or lack thereof) in the CDF-S-subtracted case. Conversely, the local field-subtracted density estimates aim to mitigate these biases by incorporating field-to-field variations, along with different off-axis completeness corrections for each field.~Therefore, for the subsequent analyses in this study, we present results exclusively from the completeness-corrected, local field-subtracted versions, which we consider more reliable under these conditions.

\section{Discussion}\label{sec:discussion}
In this work, we aim to investigate the role of the host environment in triggering X-ray AGN activity in a sample of nine massive high redshift \madcows\ clusters.~Our analysis from Figures \ref{fig:agn_sb} and \ref{fig:logn-logs_iso} suggests several of these clusters demonstrate mild correlations between AGN incidence and cluster X-ray surface brightness. This result is consistent with earlier studies of high redshift clusters/protoclusters that have shown elevated AGN activity \citep[e.g.,][]{lehmer_2009,digby_2010,martini_2013,umehata_2014,Alexander_2016,macuga_2019}.~Since  cluster formation is in a more active phase at high redshift, theoretical models suggest that the dense environments favor the significant accretion of the supermassive black holes (SMBHs) if the growth of galaxies and their central SMBHs are causally linked \citep{eastman_2007,silverman_2008,lehmer_2009}.~However, we caution the reader that this apparent environmental dependence could also stem from more fundamental properties of the host galaxy that can bias our perception \citep[e.g., stellar mass, $M_\star$; see][and references therein]{yang_2017,yang_2018}.~Hence, a careful analysis incorporating a larger sample of AGNs and galaxies, while controlling for these properties, is essential to investigate the accretion-environment relation.

One of the intriguing findings of this study is the indication of an overabundance of AGNs in disturbed clusters, evidenced by the positive correlation slopes between excess AGN surface density and photon asymmetry (see Figures \ref{fig:surf_den_iso}).~Specifically, for the $1.5r_{500}$ isophotes, we find that a disturbed cluster environment may promote AGN triggering at a $> 2\sigma$ level.~Several scenarios from the literature may potentially explain this phenomenon, if confirmed at higher significance in future studies: 
\begin{enumerate}
    \item Cluster-wide shocks that cross galaxies: Low to moderate redshift observations, theoretical models, and hydrodynamical simulations in recent years suggest that the passage of a cluster-merger shock can provide sufficient turbulence in the cool gas within member galaxies to trigger AGN activity for several Myr
    \citep{markevitch_2007,hwang_2009,sobral_2015,weeren_2019,stroe_2021}. Unfortunately, our observations are too shallow to detect any surface brightness discontinuity that may be attributed to shocks in the disturbed clusters.
    \item Active accretion of groups/filaments: \cite{ebeling_2019} proposed that ram pressure, possibly aided by shock waves, can induce star formation in gas-rich galaxies infalling along the filaments in the dense cosmic web around merging/young clusters. Similar processes can also trigger AGN activity since both rely heavily on the cold gas content \citep[e.g.,][]{santini_2014,gobat_2020}, which is expected to be higher at these redshifts \citep{tacconi_2010, munoz_2023}. 
    \item Galaxy interactions: Non-axisymmetric perturbations can invoke mass inflow during galaxy interactions/merging, which can be responsible for AGN triggering \cite[e.g.,][]{ellison_2011,koulouridis_2013,hopkins_2014}.~However, when we visually inspect optical images of these clusters, we do not see any evidence of galaxy-galaxy interactions. 
\end{enumerate}

In any event, this potential enhanced AGN triggering in dense cluster environments inspires us to explore further the dependence of excess densities on properties such as cluster mass and redshift. Our narrow redshift range ($0.819 \leqslant z \leqslant 1.230$), however, is too limited to adequately explore a redshift dependence, so we only discuss a possible mass dependence below. 

\subsection{Dependence on Cluster Mass} \label{sec:mass_dependence}
Several studies in the past reported that AGN space density ($\Phi_{\textup{AGN}}$) depends inversely on cluster mass, scaling as $M^{-1}$ \citep[e.g.,][]{ehlert_2015,noordeh_2020}. The most common interpretation of this observed behavior comes from virial arguments, where the velocity dispersion ($\sigma$) of galaxies in clusters scales with cluster mass as $M^{1/3}$, and from theoretical modeling, the galaxy merger rate in a cluster environment scales as $\sigma^{-3}$ \citep{mamon_1992}. This translates to the $M^{-1}$ scaling if galaxy-galaxy mergers and tidal interactions are the main driving mechanisms for the elevated AGN activity in clusters. 

\begin{figure}[htbp]
\includegraphics[width=\columnwidth]{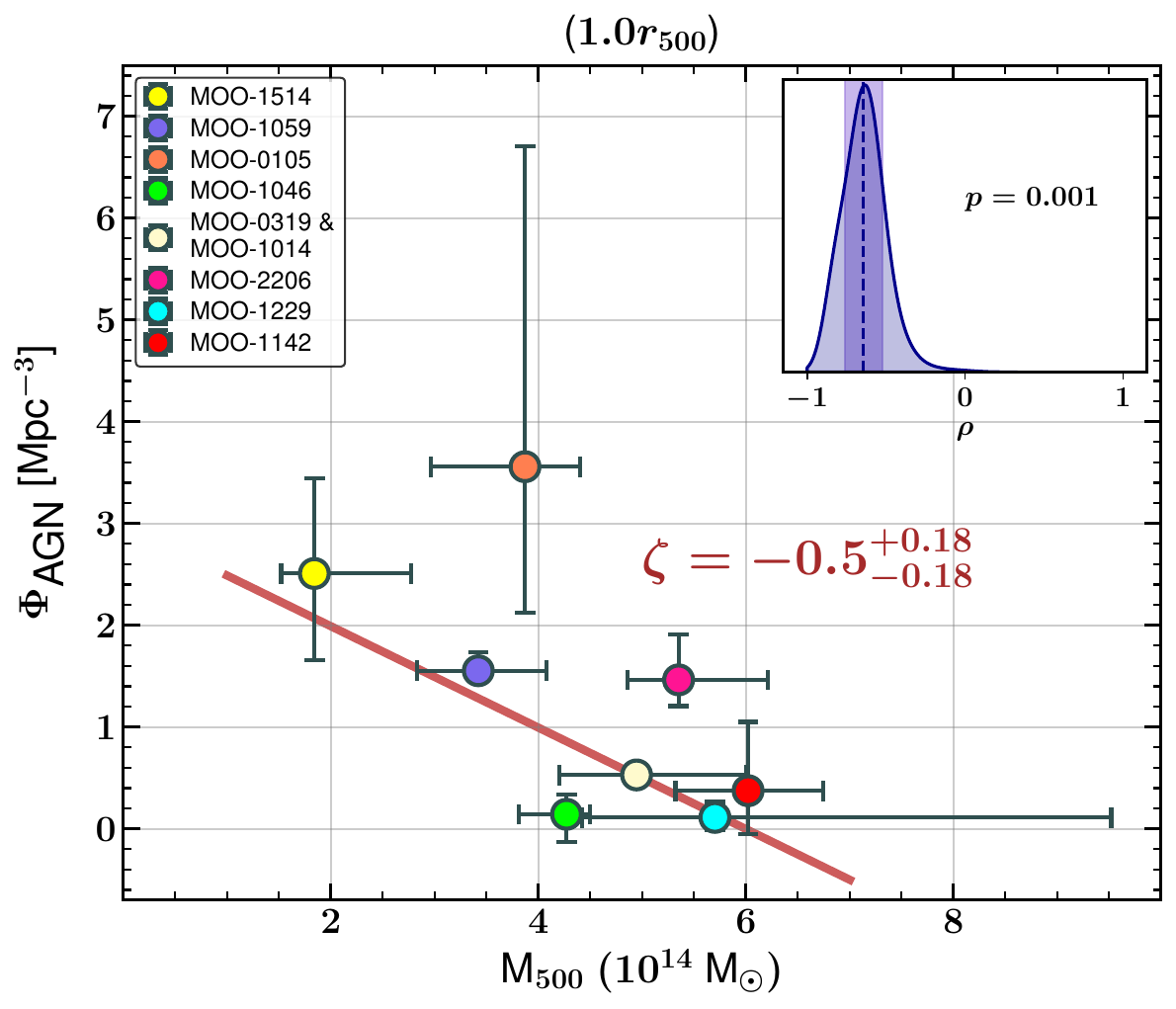}
\centering
\caption{Completeness-corrected excess AGN space density in cluster $1.0r_{500}$ isophotes as a function of cluster mass, where we subtract the field expectations estimated from the local background; hence negative values are possible. To calculate densities, we use the same luminosity cut as in Figure \ref{fig:surf_den_iso}, except we combine MOO J1014+0038 with MOO J0319--0025 and substitute spherical volumes for surface areas. As described in Section \ref{sec:mass_dependence}, we show the best fit of the power law model $M^{\zeta}$ in brown. \label{fig:mass_space}}
\end{figure}

To test this apparent dependence in our \madcows\ cluster sample, we assess the excess AGN space densities within $1.0r_{500}$ isophotes, as depicted in Figure \ref{fig:mass_space}. We employ a similar approach as described in equation \ref{eq:4}, with the exception that we replace the surface areas with spherical volumes.~Additionally, instead of binning MOO J1014+0038 solely by $A_{\text{phot}}$, we combine it with MOO J0319--0025, which has a comparable mass, a similar flux limit, and an $A_{\text{phot}}$ value that is relatively close.~Following the literature, we model the excess space density as a power law in cluster mass (i.e., $\Phi_{\textup{AGN}} \propto M^{\zeta}$), which has a best-fit $\zeta=-0.5^{+0.18}_{-0.18}$, as Figure \ref{fig:mass_space} illustrates.~This is slightly flatter compared to the $\sim M^{-2.0^{+0.8}_{-0.9}}$ and $\sim M^{-1.2^{+0.7}_{-0.7}}$ scalings found by \cite{noordeh_2020} and \citep{ehlert_2015}, respectively. The caveat, as mentioned earlier, is that our optical images do not provide any clues regarding galaxy-galaxy interactions/mergers.~However, Spearman's $\rho$ in the inset of Figure \ref{fig:mass_space} rules out the zero mass dependence ($\zeta=0$) at the $3.18\sigma$ level. This implies that adding alternative physical explanations to a pure merger-driven scenario might better model the data.

\vspace{-5pt}
\subsection{Mass Dependency on Cluster Morphology} \label{sec:mass_morph}
\begin{figure*}[htbp]
\includegraphics[scale=0.55]{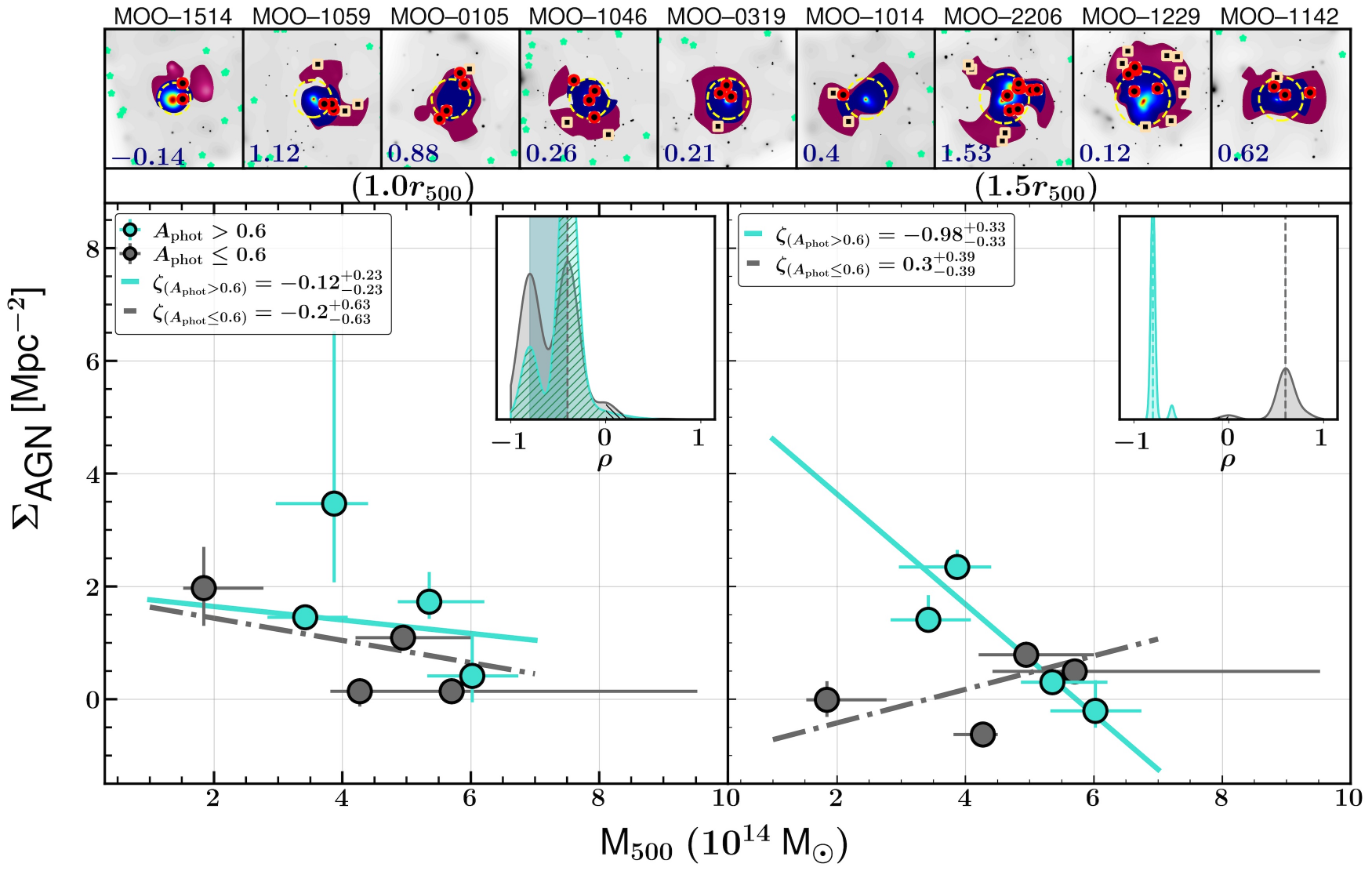}
\centering
\caption{Completeness-corrected excess AGN surface density in cluster isophotes (\textbf{left:} $1.0r_{500}$, \textbf{right:} $1.5r_{500}$) as a function of cluster mass, where we subtract the field expectations estimated from the local background; hence negative values are possible. We color-code the disturbed clusters in cyan and the relaxed clusters in grey, with the best-fit power-laws ($M^{\zeta}$) in solid and dash-dotted lines, respectively. Insets in both panels are the KDEs of Spearman’s rank correlation coefficients for the two distinct populations (i.e., disturbed vs. relaxed) matched by their corresponding colors. As before, the shaded columns represent the $1\sigma$ confidence regions around the KDEs, and the hatched regions highlight the areas corresponding to the p-values for rejecting the null hypothesis in each distribution.~The upper square tiles illustrate the cluster mass-ranked isophotes (with $A_{\text{phot}}$ values reported in the bottom-left corner of each) and their corresponding AGNs from Figure \ref{fig:aphot_masked}. \label{fig:mass_split}}
\end{figure*}

An intriguing aspect of this tentative mass dependence is apparent once we fit separate models to the two distinct cluster morphologies, i.e.\ non-merging/relaxed and merging/disturbed. As discussed earlier, $A_{\text{phot}}>0.6$ is a reasonable threshold for classifying a cluster as having a disturbed morphology for the \madcows\ sample. We color-code these in cyan as opposed to the relaxed clusters which are in grey in Figure \ref{fig:mass_split}.~For simplicity, we use $1.0r_{500}$ and $1.5r_{500}$ isophote excess surface densities here, although we can confirm that it does not materially alter the end results if one uses excess space densities instead.

The left panel of Figure \ref{fig:mass_split} suggests that the two populations may share similar power-law fits, characterized by shallow negative or near-zero slopes.~The Spearman's $\rho$ in the inset indicates a significant overlap between the two distributions.~However, the scenario changes dramatically when the cluster outskirts ($1.5r_{500}$ isophotes) are included. In this case, disturbed clusters now tend to follow an $\sim M^{-1}$ scaling ($\zeta_{(A_{\text{phot}}>0.6)}=-0.98^{+0.33}_{-0.33}$), whereas relaxed morphologies exhibit a positive or near-zero trend ($\zeta_{(A_{\text{phot}}<0.6)}=0.3^{+0.39}_{-0.39}$), as illustrated in the right panel of Figure \ref{fig:mass_split}.~The Spearman's $\rho$ distributions indicate that, indeed, the two populations differ significantly when the cluster outskirts were taken into account. Given the small sample sizes, we do not report p-values in this figure, as they become unreliable for such limited datasets.

Models and hydrodynamical simulations predict that moderate ram pressure from tenuous ICM (generally in cluster outskirts) can be responsible for angular momentum loss in the cold gas clouds in the infalling galaxies \citep{tonnesen_2009} and can cause gravitational instability in the galactic disks \citep{schulz_2001}. This ultimately ends up depositing gas onto the nuclear SMBH, leading to AGN triggering \citep{marshal_2018, ricarte_2020}. It is possible, therefore, that in the low-mass relaxed \madcows\ clusters, the ram pressure-induced AGN triggering remains inefficient in the infall regions; but becomes more important in the more massive clusters. Indeed, earlier studies have reported a high occurrence of AGN activity in the outskirts of massive, relaxed clusters \citep{ruderman_2005,fassbender_2012}. Things are slightly more complicated for the disturbed morphology cases since the environment is constantly changing, especially in the outskirts. On the other hand, \cite{koulouridis_2019} report excess AGN in the outskirts ($2.0r_{500}$--$2.5r_{500}$) of massive, dynamically disturbed clusters, so several factors may drive the apparent $M^{-1}$ scaling relation. For example, apart from cluster-merger shock-induced AGN triggering, numerical simulations show that at high redshift, cosmic filaments can funnel streams of cold gases into the cluster potential well \citep{keres_2005,dekel_2006, keres_2009}. These filaments can effectively act as shields against the ram pressure stripping/strangulation from the ICM, thus maintaining the gas reservoirs and delaying AGN quenching \citep{kotecha_2022}. Active accretion of groups can also introduce dynamically evolved new AGN to a cluster which could elevates AGN densities in the outskirts. 

Due to our small sample sizes, we cannot further test the reliability of these observed trends; however, it is evident from the narrow KDEs of Spearman's $\rho$ in the inset that AGN triggering in relaxed and disturbed morphologies behaves quite differently in the cluster outskirts. The result remains consistent even if we chose circular annuli instead of isophotes, out to at least $2.0r_{500}$ outer radii.

\section{Summary}
In this study, we have analyzed cluster morphology and X-ray AGN population in nine massive \madcows\ clusters at $z\sim1$  using \chandra\ observations.~We employed a  luminosity cut of $7.6\times10^{42}$ erg/s to probe AGNs all across our cluster fields.~Using photon asymmetry, we classified the clusters based on their X-ray emission.~We predicted excess AGN surface densities in cluster isophotes using the extrapolated $\log N- \log S$ models at this luminosity.~In this work, our focus was specifically on unveiling the influence of cluster environment on AGN activity.~Below, we summarize the most notable results:
\begin{enumerate}
    \item We find compelling evidence that the four most dynamically disturbed ($A_{\text{phot}}>0.6$) \madcows\ clusters in our sample are mergers.~We identify MOO J1142+1527 as an ongoing merger (in agreement with \citealt{ruppin_2020}) and both MOO J0105+1323 and MOO J2206+0906 as early-stage major mergers.~Similarly, we suspect MOO J1059+5454 is a major merger that has occurred within the past $\sim 1-2$ Gyr.
    \item Several clusters in our sample exhibit mild positive correlations between cluster surface brightness and AGN incidence, suggesting that AGNs may preferentially reside within the dense environment of their ICM. Furthermore, we find that the distributions of these AGNs likely trace the structure of the cluster ICMs. 
    \item Our study shows evidence at a $> 2\sigma$ level of correlation between AGN overabundance and photon asymmetry, which indicates that a merger/disturbed cluster environment may promote AGN triggering. 
    \item Overall, we find excess AGN density inversely depends on cluster mass within the $1.0r_{500}$ isophotes, following an $M^{-0.5^{+0.18}_{-0.18}}$ scaling relation at a level of $3.18\sigma$.
    \item When probing the dependence of excess AGN surface density on cluster mass in disturbed vs. relaxed cluster environments, we find tantalizing evidence that the two morphologically distinct populations exhibit dissimilar trends, potentially originating from the cluster outskirts.~This behavior may be attributed to merger-driven AGN triggering or suppression.~However, our limited sample size of clusters restricts our ability to thoroughly investigate this apparent contrasting behavior in finer detail.
 \end{enumerate}
This pilot study enables us to compare AGN populations in some of the most massive and distant clusters with the well-studied systems in the local Universe. Our results thus lend observational support to a scenario where a disturbed cluster environment stimulates AGN triggering, likely due to the cluster-merger shocks that cross galaxies, and ram pressure-driven turbulence in the infalling galaxies.~It is crucially important to consider factors such as the time elapsed since the initial core passage, collision geometry, and the orientation of the merger axis \citep{ebeling_2019}.~We acknowledge that our study potentially suffers from small number statistics, primarily attributed to the limited sensitivity of current X-ray telescopes at high redshifts. Additionally, the X-ray selection process may exhibit reduced sensitivity to Compton-thick AGNs, which can introduce bias, as highlighted by \cite{li_2019}. Thus, future large-scale studies comprising multiwavelength observations enabling unbiased selection of AGN, preferable with spectroscopic redshifts, will be more decisive in assessing our findings and conclusions.

\section*{acknowledgments}
We thank Ben Floyd for providing the Python script containing the function that calculates the upper and lower Poisson confidence limits for small number samples using \cite{gehrels_1986}.~We sincerely appreciate the valuable feedback provided by Brett Lehmer, Neil Brandt, and Ripon Saha, which significantly enhanced the quality of this article. This paper is based on observations collected at the \chandra\ X-ray Center, which is operated by the Smithsonian Astrophysical Observatory for and on behalf of NASA under contract NAS8-03060.~This publication makes use of data products from the \textit{Wide-field Infrared Survey Explorer}, which is a joint project of the University of California, Los Angeles, and the Jet Propulsion Laboratory/California Institute of Technology, funded by NASA. The work of T.C., P.R.M.E., and D.S. was carried out at the Jet Propulsion Laboratory, California Institute of Technology, under a contract with NASA. 

\facility{Chandra}
\software{CIAO \citep{ciao}, MARX \citep{marx}, emcee \citep{foreman_2013}, Astropy \citep{astropy}, 
SciPy \citep{scipy}, Matplotlib \citep{matplotlib}, NumPy \citep{numpy}, seaborn \citep{seaborn}, pandas \citep{pandas}.}

\bibliography{references}{}
\bibliographystyle{yahapj}

\end{document}

%% file: table1.tex
\begin{deluxetable*}{cccccCCCCC}[htb]
\tablenum{1}
\tablecaption{A Summary of the Cluster Sample and \chandra\ Observations 
\label{tab:data}}
\tablewidth{1pt}
\tablehead{
\colhead{Cluster ID} & \colhead{Short Name \tablenotemark{d}} & \colhead{RA} & \colhead{Dec.} & \colhead{Instrument} & \colhead{$z$} & \colhead{Exposure} & \colhead{M$_{500}$} & \colhead{$r_{500}$} & \colhead{$N_{\textup{H}}$}\\ 
\colhead{} & \colhead{} & \multicolumn{2}{c}{(J2000)} & \colhead{} & \colhead{} & \colhead{(ks)} & \colhead{($10^{14}$ \(\textup{M}_\odot\))} & \colhead{(Mpc)} & \colhead{$10^{20}$cm$^{-2}$}
}
\startdata
MOO J0105+1323 & MOO--0105 & 01:05:31.5 & +13:23:55 & ACIS-I & 1.143 & 24.21 & 3.87_{-0.91}^{+0.53} & 0.73_{-0.06}^{+0.03} & 4.20\\
MOO J0319$-$0025 & MOO--0319 & 03:19:24.4 & $-$00:25:21 & ACIS-I & 1.194 & 69.38\tablenotemark{c} & 4.81_{-1.10}^{+1.22} & 0.77_{-0.06}^{+0.07} & 6.27\\
MOO J1014+0038 & MOO--1014 & 10:14:08.4 & +00:38:26 & ACIS-I & 1.230 & 29.69 &  5.08_{-1.00}^{+1.72} & 0.77_{-0.05}^{+0.09} & 3.64\\
MOO J1046+2758 & MOO--1046 & 10:46:51.6 & +27:58:05 & ACIS-I & 1.16\tablenotemark{b} & 142.98\tablenotemark{c} &  4.27_{-0.46}^{+0.23} & 0.75_{-0.03}^{+0.01} & 2.34\\
MOO J1059+5454 & MOO--1059 & 10:59:49.4 & +54:54:37 & ACIS-I & 1.14\tablenotemark{b} & 104.57\tablenotemark{c} &  3.42_{-0.59}^{+0.66} & 0.70_{-0.04}^{+0.05} & 0.79\\
MOO J1142+1527 & MOO--1142 & 11:42:45.1 & +15:27:05 & ACIS-I & 1.189 & 46.95 & 6.02_{-0.70}^{+0.72} & 0.83_{-0.03}^{+0.03} & 2.78\\
MOO J1229+6521\tablenotemark{a} & MOO--1229 & 12:29:44.6 & +65:21:22 & ACIS-S & 0.819 & 90.01\tablenotemark{c} & 5.70_{-1.28}^{+3.82} & 0.94_{-0.07}^{+0.21} & 2.03\\
MOO J1514+1346 & MOO--1514 & 15:14:42.7 & +13:46:31 & ACIS-I & 1.059 & 70.79\tablenotemark{c} & 1.84_{-0.32}^{+0.93} & 0.59_{-0.03}^{+0.10} & 2.76\\
MOO J2206+0906 & MOO--2206 & 22:06:28.6 & +09:06:32 & ACIS-I & 0.926 & 95.50\tablenotemark{c} & 5.35_{-0.49}^{+0.86} & 0.89_{-0.03}^{+0.05} & 5.74\\
\enddata
\tablecomments{The coordinates correspond to the center of the galaxy overdensities from the \madcows\ survey \citep{gonzalez_2019}. Redshifts are from \citet{gonzalez_2019} and are spectroscopic unless otherwise specified. The exposure times (total) only include good time intervals. All mass measurements are derived from X-ray emission, based on the ICM temperature. The last column provides the intervening galactic hydrogen column density in the direction of the cluster, expressed in units of $10^{20}$ atoms cm$^{-2}$. 
\tablenotetext{a}{Also identified as PSZ2 G126.57+51.61 in the Planck survey \citep{planck_2015}.}
\tablenotetext{b}{Photometric redshifts.}
\tablenotetext{c}{Clusters with multiple observations.}
\tablenotetext{d}{Cluster short names that we use in figures.
}
}

\end{deluxetable*}

%% file: table2.tex
\begin{deluxetable}{cccc}[htb!]
\tablenum{2}
\tablecaption{Flux Limits and Completeness Levels 
\label{tab:completeness}}
\tablewidth{\columnwidth}
\tablehead{
\colhead{Cluster} & \colhead{Flux Limit} & \colhead{$C_{\leqslant3r_{500}}(\%)$} & \colhead{$C_{>3r_{500}}(\%)$}\\
\colhead{(1)} & \colhead{(2)} & 
\colhead{(3)} & \colhead{(4)}
}
\startdata
MOO J0105+1323 & 6.30 & 89 & 68\\
MOO J0319$-$0025 & 3.28 & 98 & 69\\
MOO J1014+0038 & 5.85 & 95 & 76\\
MOO J1046+2758 & 1.78 & 100 & 65\\
MOO J1059+5454 & 2.33 & 99 & 86\\
MOO J1142+1527 & 4.29 & 94 & 84\\
MOO J1229+6521 & 2.18 & 100 & 90\\
MOO J1514+1346 & 3.44 & 93 & 89\\
MOO J2206+0906 & 2.41 & 100 & 54\\
\enddata
\tablecomments{Column~2:~Cluster-specific flux limits in the 0.7--7.0 keV band, representing the minimum flux at which 50\% of that field is sensitive, expressed in units of $10^{-15}$ erg cm$^{-2}$ s$^{-1}$.~Columns~3-4: X-ray point source detection completeness levels within the cluster ($C_{\leqslant3r_{500}}$) and local field ($C_{>3r_{500}}$) regions for our survey luminosity cut of $L_{X} = 7.6\times10^{42}$ erg/s.
}
\end{deluxetable}

%% file: table3.tex
\begin{deluxetable*}{cccccc}[htb!]
\tablenum{3}
\tablecaption{Best-fit Parameters with $1\sigma$ Errors for the $dN/dS$ Model 
\label{tab:logN-logS}}
\tablewidth{\columnwidth}
\tablehead{
\colhead{AGNs} & \colhead{Bandpass} & \colhead{$K$} & \colhead{ $\beta_{1}$} & \colhead{ $\beta_{2}$} & \colhead{$f_{b}$}\\
\colhead{(1)} & \colhead{(2)} & \colhead{(3)} & \colhead{(4)} & \colhead{(5)} & \colhead{(6)}
}
\startdata
MaDCoWS $(\leq3r_{500})$ & 
$0.7-7.0$ keV &
$1312.34_{-483.96}^{+437.18}$ & $-0.30_{-0.29}^{+0.36}$ & 
$2.29_{-0.23}^{+0.33}$ & 
$6.2_{-0.7}^{+1.2}$\\
MaDCoWS $(>3r_{500})$ &
$0.7-7.0$ keV &
$1055.81_{-345.24}^{+547.73}$ & $-0.71_{-0.38}^{+0.38}$ & 
$2.01_{-0.13}^{+0.15}$ & 
$6.3_{-0.9}^{+1.1}$\\
CDF-S \citepalias{lehmer_2012} &
$0.5-8.0$ keV &
$562.20_{-22.96}^{+22.96}$ & $1.34_{-0.03}^{+0.04}$ & 
$2.35_{-0.15}^{+0.15}$ & 
$8.1_{-1.4}^{+1.5}$\\
\enddata

\tablecomments{Column~1: source of AGNs, Column~2: X-ray band.~Columns~3–6: normalization in units of $10^{14}$ deg$^{-2}$ [erg cm$^{-2}$ s$^{-1}$]$^{-1}$, faint-end slope, bright-end slope, and break flux in units of $10^{-15}$ erg cm$^{-2}$ s$^{-1}$ for the AGN double power-law $dN/dS$ model.
}
\end{deluxetable*}

%% file: main.bbl
\begin{thebibliography}{}
\providecommand\natexlab[1]{#1}
\providecommand\JournalTitle[1]{#1}

\bibitem[{{Alberts} {et~al.}(2016){Alberts}, {Pope}, {Brodwin}, {Chung},
  {Cybulski}, {Dey}, {Eisenhardt}, {Galametz}, {Gonzalez}, {Jannuzi},
  {Stanford}, {Snyder}, {Stern}, \& {Zeimann}}]{alberts_2016}
{Alberts}, S., {Pope}, A., {Brodwin}, M., {et~al.} 2016,
  \href{http://dx.doi.org/10.3847/0004-637X/825/1/72}{\JournalTitle{\apj}, 825,
  72}

\bibitem[{{Alexander} {et~al.}(2016){Alexander}, {Simpson}, {Harrison},
  {Mullaney}, {Smail}, {Geach}, {Hickox}, {Hine}, {Karim}, {Kubo}, {Lehmer},
  {Matsuda}, {Rosario}, {Stanley}, {Swinbank}, {Umehata}, \&
  {Yamada}}]{Alexander_2016}
{Alexander}, D.~M., {Simpson}, J.~M., {Harrison}, C.~M., {et~al.} 2016,
  \href{http://dx.doi.org/10.1093/mnras/stw1509}{\JournalTitle{\mnras}, 461,
  2944}

\bibitem[{{Andersson} {et~al.}(2011){Andersson}, {Benson}, {Ade}, {Aird},
  {Armstrong}, {Bautz}, {Bleem}, {Brodwin}, {Carlstrom}, {Chang}, {Crawford},
  {Crites}, {de Haan}, {Desai}, {Dobbs}, {Dudley}, {Foley}, {Forman},
  {Garmire}, {George}, {Gladders}, {Halverson}, {High}, {Holder}, {Holzapfel},
  {Hrubes}, {Jones}, {Joy}, {Keisler}, {Knox}, {Lee}, {Leitch}, {Lueker},
  {Marrone}, {McMahon}, {Mehl}, {Meyer}, {Mohr}, {Montroy}, {Murray}, {Padin},
  {Plagge}, {Pryke}, {Reichardt}, {Rest}, {Ruel}, {Ruhl}, {Schaffer}, {Shaw},
  {Shirokoff}, {Song}, {Spieler}, {Stalder}, {Staniszewski}, {Stark}, {Stubbs},
  {Vanderlinde}, {Vieira}, {Vikhlinin}, {Williamson}, {Yang}, {Zahn}, \&
  {Zenteno}}]{andersson_2011}
{Andersson}, K., {Benson}, B.~A., {Ade}, P.~A.~R., {et~al.} 2011,
  \href{http://dx.doi.org/10.1088/0004-637X/738/1/48}{\JournalTitle{\apj}, 738,
  48}

\bibitem[{{Astropy Collaboration} {et~al.}(2013){Astropy Collaboration},
  {Robitaille}, {Tollerud}, {Greenfield}, {Droettboom}, {Bray}, {Aldcroft},
  {Davis}, {Ginsburg}, {Price-Whelan}, {Kerzendorf}, {Conley}, {Crighton},
  {Barbary}, {Muna}, {Ferguson}, {Grollier}, {Parikh}, {Nair}, {Unther},
  {Deil}, {Woillez}, {Conseil}, {Kramer}, {Turner}, {Singer}, {Fox}, {Weaver},
  {Zabalza}, {Edwards}, {Azalee Bostroem}, {Burke}, {Casey}, {Crawford},
  {Dencheva}, {Ely}, {Jenness}, {Labrie}, {Lim}, {Pierfederici}, {Pontzen},
  {Ptak}, {Refsdal}, {Servillat}, \& {Streicher}}]{astropy}
{Astropy Collaboration}, {Robitaille}, T.~P., {Tollerud}, E.~J., {et~al.} 2013,
  \href{http://dx.doi.org/10.1051/0004-6361/201322068}{\JournalTitle{\aap},
  558, A33}

\bibitem[{{Balmaverde} {et~al.}(2017){Balmaverde}, {Gilli}, {Mignoli},
  {Bolzonella}, {Brusa}, {Cappelluti}, {Comastri}, {Sani}, {Vanzella},
  {Vignali}, {Vito}, \& {Zamorani}}]{balmaverde_2017}
{Balmaverde}, B., {Gilli}, R., {Mignoli}, M., {et~al.} 2017,
  \href{http://dx.doi.org/10.1051/0004-6361/201730683}{\JournalTitle{\aap},
  606, A23}

\bibitem[{{Balogh} {et~al.}(2004){Balogh}, {Eke}, {Miller}, {Lewis}, {Bower},
  {Couch}, {Nichol}, {Bland-Hawthorn}, {Baldry}, {Baugh}, {Bridges}, {Cannon},
  {Cole}, {Colless}, {Collins}, {Cross}, {Dalton}, {de Propris}, {Driver},
  {Efstathiou}, {Ellis}, {Frenk}, {Glazebrook}, {Gomez}, {Gray}, {Hawkins},
  {Jackson}, {Lahav}, {Lumsden}, {Maddox}, {Madgwick}, {Norberg}, {Peacock},
  {Percival}, {Peterson}, {Sutherland}, \& {Taylor}}]{balogh_2004}
{Balogh}, M., {Eke}, V., {Miller}, C., {et~al.} 2004,
  \href{http://dx.doi.org/10.1111/j.1365-2966.2004.07453.x}{\JournalTitle{\mnras},
  348, 1355}

\bibitem[{{Bauer} {et~al.}(2004){Bauer}, {Alexander}, {Brandt}, {Schneider},
  {Treister}, {Hornschemeier}, \& {Garmire}}]{bauer_2004A}
{Bauer}, F.~E., {Alexander}, D.~M., {Brandt}, W.~N., {et~al.} 2004,
  \href{http://dx.doi.org/10.1086/424859}{\JournalTitle{\aj}, 128, 2048}

\bibitem[{{Best}(2004)}]{best_2004}
{Best}, P.~N. 2004,
  \href{http://dx.doi.org/10.1111/j.1365-2966.2004.07752.x}{\JournalTitle{\mnras},
  351, 70}

\bibitem[{{B{\"o}hringer} {et~al.}(2010){B{\"o}hringer}, {Pratt}, {Arnaud},
  {Borgani}, {Croston}, {Ponman}, {Ameglio}, {Temple}, \&
  {Dolag}}]{bohringer_2010}
{B{\"o}hringer}, H., {Pratt}, G.~W., {Arnaud}, M., {et~al.} 2010,
  \href{http://dx.doi.org/10.1051/0004-6361/200913911}{\JournalTitle{\aap},
  514, A32}

\bibitem[{{Brandt} \& {Alexander}(2015)}]{brandt_2015}
{Brandt}, W.~N., \& {Alexander}, D.~M. 2015,
  \href{http://dx.doi.org/10.1007/s00159-014-0081-z}{\JournalTitle{\aapr}, 23,
  1}

\bibitem[{{Brodwin} {et~al.}(2015){Brodwin}, {Greer}, {Leitch}, {Stanford},
  {Gonzalez}, {Gettings}, {Abdulla}, {Carlstrom}, {Decker}, {Eisenhardt},
  {Lin}, {Mantz}, {Marrone}, {McDonald}, {Stalder}, {Stern}, \&
  {Wylezalek}}]{brodwin_2015}
{Brodwin}, M., {Greer}, C.~H., {Leitch}, E.~M., {et~al.} 2015,
  \href{http://dx.doi.org/10.1088/0004-637X/806/1/26}{\JournalTitle{\apj}, 806,
  26}

\bibitem[{{Bufanda} {et~al.}(2017){Bufanda}, {Hollowood}, {Jeltema}, {Rykoff},
  {Rozo}, {Martini}, {Abbott}, {Abdalla}, {Allam}, {Banerji},
  {Benoit-L{\'e}vy}, {Bertin}, {Brooks}, {Carnero Rosell}, {Carrasco Kind},
  {Carretero}, {Cunha}, {da Costa}, {Desai}, {Diehl}, {Dietrich}, {Evrard},
  {Fausti Neto}, {Flaugher}, {Frieman}, {Gerdes}, {Goldstein}, {Gruen},
  {Gruendl}, {Gutierrez}, {Honscheid}, {James}, {Kuehn}, {Kuropatkin}, {Lima},
  {Maia}, {Marshall}, {Melchior}, {Miquel}, {Mohr}, {Ogando}, {Plazas},
  {Romer}, {Rooney}, {Sanchez}, {Santiago}, {Scarpine}, {Sevilla-Noarbe},
  {Smith}, {Soares-Santos}, {Sobreira}, {Suchyta}, {Tarle}, {Thomas}, {Tucker},
  {Walker}, \& {DES Collaboration}}]{bufanda_2017}
{Bufanda}, E., {Hollowood}, D., {Jeltema}, T.~E., {et~al.} 2017,
  \href{http://dx.doi.org/10.1093/mnras/stw2824}{\JournalTitle{\mnras}, 465,
  2531}

\bibitem[{{Buote} \& {Tsai}(1995)}]{buote_1995}
{Buote}, D.~A., \& {Tsai}, J.~C. 1995,
  \href{http://dx.doi.org/10.1086/176326}{\JournalTitle{\apj}, 452, 522}

\bibitem[{{Buote} \& {Tsai}(1996)}]{buote_1996}
{Buote}, D.~A., \& {Tsai}, J.~C. 1996,
  \href{http://dx.doi.org/10.1086/176790}{\JournalTitle{\apj}, 458, 27}

\bibitem[{{Chon} {et~al.}(2016){Chon}, {Puchwein}, \&
  {B{\"o}hringer}}]{chong_2016}
{Chon}, G., {Puchwein}, E., \& {B{\"o}hringer}, H. 2016,
  \href{http://dx.doi.org/10.1051/0004-6361/201628532}{\JournalTitle{\aap},
  592, A46}

\bibitem[{{Cowie} \& {Songaila}(1977)}]{cowie_1977}
{Cowie}, L.~L., \& {Songaila}, A. 1977,
  \href{http://dx.doi.org/10.1038/266501a0}{\JournalTitle{\nat}, 266, 501}

\bibitem[{{Croton} {et~al.}(2006){Croton}, {Springel}, {White}, {De Lucia},
  {Frenk}, {Gao}, {Jenkins}, {Kauffmann}, {Navarro}, \&
  {Yoshida}}]{croton_2006}
{Croton}, D.~J., {Springel}, V., {White}, S. D.~M., {et~al.} 2006,
  \href{http://dx.doi.org/10.1111/j.1365-2966.2005.09675.x}{\JournalTitle{\mnras},
  365, 11}

\bibitem[{{Darvish} {et~al.}(2017){Darvish}, {Mobasher}, {Martin}, {Sobral},
  {Scoville}, {Stroe}, {Hemmati}, \& {Kartaltepe}}]{darvish_2017}
{Darvish}, B., {Mobasher}, B., {Martin}, D.~C., {et~al.} 2017,
  \href{http://dx.doi.org/10.3847/1538-4357/837/1/16}{\JournalTitle{\apj}, 837,
  16}

\bibitem[{{Davidzon} {et~al.}(2016){Davidzon}, {Cucciati}, {Bolzonella}, {De
  Lucia}, {Zamorani}, {Arnouts}, {Moutard}, {Ilbert}, {Garilli}, {Scodeggio},
  {Guzzo}, {Abbas}, {Adami}, {Bel}, {Bottini}, {Branchini}, {Cappi}, {Coupon},
  {de la Torre}, {Di Porto}, {Fritz}, {Franzetti}, {Fumana}, {Granett},
  {Guennou}, {Iovino}, {Krywult}, {Le Brun}, {Le F{\`e}vre}, {Maccagni},
  {Ma{\l}ek}, {Marulli}, {McCracken}, {Mellier}, {Moscardini}, {Polletta},
  {Pollo}, {Tasca}, {Tojeiro}, {Vergani}, \& {Zanichelli}}]{davidzon_2016}
{Davidzon}, I., {Cucciati}, O., {Bolzonella}, M., {et~al.} 2016,
  \href{http://dx.doi.org/10.1051/0004-6361/201527129}{\JournalTitle{\aap},
  586, A23}

\bibitem[{{Davis} {et~al.}(2012){Davis}, {Bautz}, {Dewey}, {Heilmann}, {Houck},
  {Huenemoerder}, {Marshall}, {Nowak}, {Schattenburg}, {Schulz}, \&
  {Smith}}]{marx}
{Davis}, J.~E., {Bautz}, M.~W., {Dewey}, D., {et~al.} 2012,
  \href{http://dx.doi.org/10.1117/12.926937}{in Society of Photo-Optical
  Instrumentation Engineers (SPIE) Conference Series, Vol. 8443, Space
  Telescopes and Instrumentation 2012: Ultraviolet to Gamma Ray, ed.
  T.~{Takahashi}, S.~S. {Murray}, \& J.-W.~A. {den Herder}}, 84431A

\bibitem[{{Decker} {et~al.}(2022){Decker}, {Brodwin}, {Saha}, {Connor},
  {Eisenhardt}, {Gonzalez}, {Moravec}, {Muhibullah}, {Stanford}, {Stern},
  {Thongkham}, {Wylezalek}, {Dicker}, {Mason}, {Mroczkowski}, {Romero}, \&
  {Ruppin}}]{decker_2022}
{Decker}, B., {Brodwin}, M., {Saha}, R., {et~al.} 2022,
  \href{http://dx.doi.org/10.3847/1538-4357/ac85e5}{\JournalTitle{\apj}, 936,
  71}

\bibitem[{{Dekel} \& {Birnboim}(2006)}]{dekel_2006}
{Dekel}, A., \& {Birnboim}, Y. 2006,
  \href{http://dx.doi.org/10.1111/j.1365-2966.2006.10145.x}{\JournalTitle{\mnras},
  368, 2}

\bibitem[{{Dicker} {et~al.}(2020){Dicker}, {Romero}, {Di Mascolo},
  {Mroczkowski}, {Sievers}, {Moravec}, {Bhandarkar}, {Brodwin}, {Connor},
  {Decker}, {Devlin}, {Gonzalez}, {Lowe}, {Mason}, {Sarazin}, {Stanford},
  {Stern}, {Thongkham}, {Wylezalek}, \& {Zago}}]{dicker_2020}
{Dicker}, S.~R., {Romero}, C.~E., {Di Mascolo}, L., {et~al.} 2020,
  \href{http://dx.doi.org/10.3847/1538-4357/abb673}{\JournalTitle{\apj}, 902,
  144}

\bibitem[{{Dickey} {et~al.}(2016){Dickey}, {van Dokkum}, {Oesch}, {Whitaker},
  {Momcheva}, {Nelson}, {Leja}, {Brammer}, {Franx}, \& {Skelton}}]{dickey_2016}
{Dickey}, C.~M., {van Dokkum}, P.~G., {Oesch}, P.~A., {et~al.} 2016,
  \href{http://dx.doi.org/10.3847/2041-8205/828/1/L11}{\JournalTitle{\apjl},
  828, L11}

\bibitem[{{Digby-North} {et~al.}(2010){Digby-North}, {Nandra}, {Laird},
  {Steidel}, {Georgakakis}, {Bogosavljevi{\'c}}, {Erb}, {Shapley}, {Reddy}, \&
  {Aird}}]{digby_2010}
{Digby-North}, J.~A., {Nandra}, K., {Laird}, E.~S., {et~al.} 2010,
  \href{http://dx.doi.org/10.1111/j.1365-2966.2010.16977.x}{\JournalTitle{\mnras},
  407, 846}

\bibitem[{{Dressler}(1980)}]{dressler_1980}
{Dressler}, A. 1980,
  \href{http://dx.doi.org/10.1086/157753}{\JournalTitle{\apj}, 236, 351}

\bibitem[{{Dressler} {et~al.}(1985){Dressler}, {Thompson}, \&
  {Shectman}}]{dressler_1985}
{Dressler}, A., {Thompson}, I.~B., \& {Shectman}, S.~A. 1985,
  \href{http://dx.doi.org/10.1086/162813}{\JournalTitle{\apj}, 288, 481}

\bibitem[{{Eastman} {et~al.}(2007){Eastman}, {Martini}, {Sivakoff}, {Kelson},
  {Mulchaey}, \& {Tran}}]{eastman_2007}
{Eastman}, J., {Martini}, P., {Sivakoff}, G., {et~al.} 2007,
  \href{http://dx.doi.org/10.1086/520577}{\JournalTitle{\apjl}, 664, L9}

\bibitem[{{Ebeling} \& {Kalita}(2019)}]{ebeling_2019}
{Ebeling}, H., \& {Kalita}, B.~S. 2019,
  \href{http://dx.doi.org/10.3847/1538-4357/ab35d6}{\JournalTitle{\apj}, 882,
  127}

\bibitem[{{Ebeling} {et~al.}(2014){Ebeling}, {Stephenson}, \&
  {Edge}}]{ebeling_2014}
{Ebeling}, H., {Stephenson}, L.~N., \& {Edge}, A.~C. 2014,
  \href{http://dx.doi.org/10.1088/2041-8205/781/2/L40}{\JournalTitle{\apjl},
  781, L40}

\bibitem[{{Ehlert} {et~al.}(2013){Ehlert}, {Allen}, {Brandt}, {Xue}, {Luo},
  {von der Linden}, {Mantz}, \& {Morris}}]{ehlert_2013}
{Ehlert}, S., {Allen}, S.~W., {Brandt}, W.~N., {et~al.} 2013,
  \href{http://dx.doi.org/10.1093/mnras/sts288}{\JournalTitle{\mnras}, 428,
  3509}

\bibitem[{{Ehlert} {et~al.}(2014){Ehlert}, {von der Linden}, {Allen}, {Brandt},
  {Xue}, {Luo}, {Mantz}, {Morris}, {Applegate}, \& {Kelly}}]{ehlert_2014}
{Ehlert}, S., {von der Linden}, A., {Allen}, S.~W., {et~al.} 2014,
  \href{http://dx.doi.org/10.1093/mnras/stt2025}{\JournalTitle{\mnras}, 437,
  1942}

\bibitem[{{Ehlert} {et~al.}(2015){Ehlert}, {Allen}, {Brandt}, {Canning}, {Luo},
  {Mantz}, {Morris}, {von der Linden}, \& {Xue}}]{ehlert_2015}
{Ehlert}, S., {Allen}, S.~W., {Brandt}, W.~N., {et~al.} 2015,
  \href{http://dx.doi.org/10.1093/mnras/stu2091}{\JournalTitle{\mnras}, 446,
  2709}

\bibitem[{{Ellison} {et~al.}(2011){Ellison}, {Patton}, {Mendel}, \&
  {Scudder}}]{ellison_2011}
{Ellison}, S.~L., {Patton}, D.~R., {Mendel}, J.~T., \& {Scudder}, J.~M. 2011,
  \href{http://dx.doi.org/10.1111/j.1365-2966.2011.19624.x}{\JournalTitle{\mnras},
  418, 2043}

\bibitem[{{Farouki} \& {Shapiro}(1981)}]{farouki_1981}
{Farouki}, R., \& {Shapiro}, S.~L. 1981,
  \href{http://dx.doi.org/10.1086/158563}{\JournalTitle{\apj}, 243, 32}

\bibitem[{{Fassbender} {et~al.}(2012){Fassbender}, {{\v{S}}uhada}, \&
  {Nastasi}}]{fassbender_2012}
{Fassbender}, R., {{\v{S}}uhada}, R., \& {Nastasi}, A. 2012,
  \href{http://dx.doi.org/10.1155/2012/138380}{\JournalTitle{Advances in
  Astronomy}, 2012, 138380}

\bibitem[{{Foreman-Mackey} {et~al.}(2013){Foreman-Mackey}, {Hogg}, {Lang}, \&
  {Goodman}}]{foreman_2013}
{Foreman-Mackey}, D., {Hogg}, D.~W., {Lang}, D., \& {Goodman}, J. 2013,
  \href{http://dx.doi.org/10.1086/670067}{\JournalTitle{\pasp}, 125, 306}

\bibitem[{{Freeman} {et~al.}(2002){Freeman}, {Kashyap}, {Rosner}, \&
  {Lamb}}]{freeman_2002}
{Freeman}, P.~E., {Kashyap}, V., {Rosner}, R., \& {Lamb}, D.~Q. 2002,
  \href{http://dx.doi.org/10.1086/324017}{\JournalTitle{\apjs}, 138, 185}

\bibitem[{{Fruscione} {et~al.}(2006){Fruscione}, {McDowell}, {Allen},
  {Brickhouse}, {Burke}, {Davis}, {Durham}, {Elvis}, {Galle}, {Harris},
  {Huenemoerder}, {Houck}, {Ishibashi}, {Karovska}, {Nicastro}, {Noble},
  {Nowak}, {Primini}, {Siemiginowska}, {Smith}, \& {Wise}}]{ciao}
{Fruscione}, A., {McDowell}, J.~C., {Allen}, G.~E., {et~al.} 2006,
  \href{http://dx.doi.org/10.1117/12.671760}{\JournalTitle{Proc. SPIE}, 6270,
  62701V}

\bibitem[{{Galametz} {et~al.}(2009){Galametz}, {Stern}, {Eisenhardt},
  {Brodwin}, {Brown}, {Dey}, {Gonzalez}, {Jannuzi}, {Moustakas}, \&
  {Stanford}}]{galametz_2009}
{Galametz}, A., {Stern}, D., {Eisenhardt}, P. R.~M., {et~al.} 2009,
  \href{http://dx.doi.org/10.1088/0004-637X/694/2/1309}{\JournalTitle{\apj},
  694, 1309}

\bibitem[{{Gehrels}(1986)}]{gehrels_1986}
{Gehrels}, N. 1986,
  \href{http://dx.doi.org/10.1086/164079}{\JournalTitle{\apj}, 303, 336}

\bibitem[{{Georgakakis} {et~al.}(2008){Georgakakis}, {Nandra}, {Laird}, {Aird},
  \& {Trichas}}]{Georgakakis_2008}
{Georgakakis}, A., {Nandra}, K., {Laird}, E.~S., {Aird}, J., \& {Trichas}, M.
  2008,
  \href{http://dx.doi.org/10.1111/j.1365-2966.2008.13423.x}{\JournalTitle{\mnras},
  388, 1205}

\bibitem[{{Gettings} {et~al.}(2012){Gettings}, {Gonzalez}, {Stanford},
  {Eisenhardt}, {Brodwin}, {Mancone}, {Stern}, {Zeimann}, {Masci}, {Papovich},
  {Tanaka}, \& {Wright}}]{Gettings_2012}
{Gettings}, D.~P., {Gonzalez}, A.~H., {Stanford}, S.~A., {et~al.} 2012,
  \href{http://dx.doi.org/10.1088/2041-8205/759/1/L23}{\JournalTitle{\apjl},
  759, L23}

\bibitem[{{Gilmour} {et~al.}(2009){Gilmour}, {Best}, \&
  {Almaini}}]{gilmour_2009}
{Gilmour}, R., {Best}, P., \& {Almaini}, O. 2009,
  \href{http://dx.doi.org/10.1111/j.1365-2966.2008.14161.x}{\JournalTitle{\mnras},
  392, 1509}

\bibitem[{{Gobat} {et~al.}(2020){Gobat}, {Magdis}, {D'Eugenio}, \&
  {Valentino}}]{gobat_2020}
{Gobat}, R., {Magdis}, G., {D'Eugenio}, C., \& {Valentino}, F. 2020,
  \href{http://dx.doi.org/10.1051/0004-6361/202039593}{\JournalTitle{\aap},
  644, L7}

\bibitem[{{Gonzalez} {et~al.}(2019){Gonzalez}, {Gettings}, {Brodwin},
  {Eisenhardt}, {Stanford}, {Wylezalek}, {Decker}, {Marrone}, {Moravec},
  {O'Donnell}, {Stalder}, {Stern}, {Abdulla}, {Brown}, {Carlstrom}, {Chambers},
  {Hayden}, {Lin}, {Magnier}, {Masci}, {Mantz}, {McDonald}, {Mo}, {Perlmutter},
  {Wright}, \& {Zeimann}}]{gonzalez_2019}
{Gonzalez}, A.~H., {Gettings}, D.~P., {Brodwin}, M., {et~al.} 2019,
  \href{http://dx.doi.org/10.3847/1538-4365/aafad2}{\JournalTitle{\apjs}, 240,
  33}

\bibitem[{{Green} {et~al.}(2019){Green}, {Ntampaka}, {Nagai}, {Lovisari},
  {Dolag}, {Eckert}, \& {ZuHone}}]{green_2019}
{Green}, S.~B., {Ntampaka}, M., {Nagai}, D., {et~al.} 2019,
  \href{http://dx.doi.org/10.3847/1538-4357/ab426f}{\JournalTitle{\apj}, 884,
  33}

\bibitem[{{Gunn} \& {Gott}(1972)}]{gunn_1972}
{Gunn}, J.~E., \& {Gott}, J.~Richard, I. 1972,
  \href{http://dx.doi.org/10.1086/151605}{\JournalTitle{\apj}, 176, 1}

\bibitem[{{Haggard} {et~al.}(2010){Haggard}, {Green}, {Anderson}, {Constantin},
  {Aldcroft}, {Kim}, \& {Barkhouse}}]{haggard_2010}
{Haggard}, D., {Green}, P.~J., {Anderson}, S.~F., {et~al.} 2010,
  \href{http://dx.doi.org/10.1088/0004-637X/723/2/1447}{\JournalTitle{\apj},
  723, 1447}

\bibitem[{Harris {et~al.}(2020)Harris, Millman, van~der Walt, Gommers,
  Virtanen, Cournapeau, Wieser, Taylor, Berg, Smith, Kern, Picus, Hoyer, van
  Kerkwijk, Brett, Haldane, del R{\'{i}}o, Wiebe, Peterson,
  G{\'{e}}rard-Marchant, Sheppard, Reddy, Weckesser, Abbasi, Gohlke, \&
  Oliphant}]{numpy}
Harris, C.~R., Millman, K.~J., van~der Walt, S.~J., {et~al.} 2020,
  \href{http://dx.doi.org/10.1038/s41586-020-2649-2}{\JournalTitle{Nature},
  585, 357}

\bibitem[{{Hopkins} \& {Beacom}(2006)}]{hopkins_2006}
{Hopkins}, A.~M., \& {Beacom}, J.~F. 2006,
  \href{http://dx.doi.org/10.1086/506610}{\JournalTitle{\apj}, 651, 142}

\bibitem[{{Hopkins}(2012)}]{hopkins_2012}
{Hopkins}, P.~F. 2012,
  \href{http://dx.doi.org/10.1111/j.1745-3933.2011.01179.x}{\JournalTitle{\mnras},
  420, L8}

\bibitem[{{Hopkins} {et~al.}(2014){Hopkins}, {Kocevski}, \&
  {Bundy}}]{hopkins_2014}
{Hopkins}, P.~F., {Kocevski}, D.~D., \& {Bundy}, K. 2014,
  \href{http://dx.doi.org/10.1093/mnras/stu1736}{\JournalTitle{\mnras}, 445,
  823}

\bibitem[{Hunter(2007)}]{matplotlib}
Hunter, J.~D. 2007,
  \href{http://dx.doi.org/10.1109/MCSE.2007.55}{\JournalTitle{Computing in
  Science \& Engineering}, 9, 90}

\bibitem[{{Hwang} \& {Lee}(2009)}]{hwang_2009}
{Hwang}, H.~S., \& {Lee}, M.~G. 2009,
  \href{http://dx.doi.org/10.1111/j.1365-2966.2009.15100.x}{\JournalTitle{\mnras},
  397, 2111}

\bibitem[{{Jeltema} {et~al.}(2005){Jeltema}, {Canizares}, {Bautz}, \&
  {Buote}}]{jeltema_2005}
{Jeltema}, T.~E., {Canizares}, C.~R., {Bautz}, M.~W., \& {Buote}, D.~A. 2005,
  \href{http://dx.doi.org/10.1086/428940}{\JournalTitle{\apj}, 624, 606}

\bibitem[{{Kapferer} {et~al.}(2006){Kapferer}, {Ferrari}, {Domainko}, {Mair},
  {Kronberger}, {Schindler}, {Kimeswenger}, {van Kampen}, {Breitschwerdt}, \&
  {Ruffert}}]{kapferer_2006}
{Kapferer}, W., {Ferrari}, C., {Domainko}, W., {et~al.} 2006,
  \href{http://dx.doi.org/10.1051/0004-6361:20053975}{\JournalTitle{\aap}, 447,
  827}

\bibitem[{{Kauffmann} {et~al.}(2004){Kauffmann}, {White}, {Heckman},
  {M{\'e}nard}, {Brinchmann}, {Charlot}, {Tremonti}, \&
  {Brinkmann}}]{kauffmann_2004}
{Kauffmann}, G., {White}, S. D.~M., {Heckman}, T.~M., {et~al.} 2004,
  \href{http://dx.doi.org/10.1111/j.1365-2966.2004.08117.x}{\JournalTitle{\mnras},
  353, 713}

\bibitem[{{Kere{\v{s}}} {et~al.}(2009){Kere{\v{s}}}, {Katz}, {Fardal},
  {Dav{\'e}}, \& {Weinberg}}]{keres_2009}
{Kere{\v{s}}}, D., {Katz}, N., {Fardal}, M., {Dav{\'e}}, R., \& {Weinberg},
  D.~H. 2009,
  \href{http://dx.doi.org/10.1111/j.1365-2966.2009.14541.x}{\JournalTitle{\mnras},
  395, 160}

\bibitem[{{Kere{\v{s}}} {et~al.}(2005){Kere{\v{s}}}, {Katz}, {Weinberg}, \&
  {Dav{\'e}}}]{keres_2005}
{Kere{\v{s}}}, D., {Katz}, N., {Weinberg}, D.~H., \& {Dav{\'e}}, R. 2005,
  \href{http://dx.doi.org/10.1111/j.1365-2966.2005.09451.x}{\JournalTitle{\mnras},
  363, 2}

\bibitem[{{Kodama} {et~al.}(2004){Kodama}, {Balogh}, {Smail}, {Bower}, \&
  {Nakata}}]{kodama_2004}
{Kodama}, T., {Balogh}, M.~L., {Smail}, I., {Bower}, R.~G., \& {Nakata}, F.
  2004,
  \href{http://dx.doi.org/10.1111/j.1365-2966.2004.08271.x}{\JournalTitle{\mnras},
  354, 1103}

\bibitem[{{Kotecha} {et~al.}(2022){Kotecha}, {Welker}, {Zhou}, {Wadsley},
  {Kraljic}, {Sorce}, {Rasia}, {Roberts}, {Gray}, {Yepes}, \&
  {Cui}}]{kotecha_2022}
{Kotecha}, S., {Welker}, C., {Zhou}, Z., {et~al.} 2022,
  \href{http://dx.doi.org/10.1093/mnras/stac300}{\JournalTitle{\mnras}, 512,
  926}

\bibitem[{{Koulouridis} \& {Bartalucci}(2019)}]{koulouridis_2019}
{Koulouridis}, E., \& {Bartalucci}, I. 2019,
  \href{http://dx.doi.org/10.1051/0004-6361/201935082}{\JournalTitle{\aap},
  623, L10}

\bibitem[{{Koulouridis} \& {Plionis}(2010)}]{koulouridis_2010}
{Koulouridis}, E., \& {Plionis}, M. 2010,
  \href{http://dx.doi.org/10.1088/2041-8205/714/2/L181}{\JournalTitle{\apjl},
  714, L181}

\bibitem[{{Koulouridis} {et~al.}(2013){Koulouridis}, {Plionis}, {Chavushyan},
  {Dultzin}, {Krongold}, {Georgantopoulos}, \&
  {Le{\'o}n-Tavares}}]{koulouridis_2013}
{Koulouridis}, E., {Plionis}, M., {Chavushyan}, V., {et~al.} 2013,
  \href{http://dx.doi.org/10.1051/0004-6361/201219606}{\JournalTitle{\aap},
  552, A135}

\bibitem[{{Koulouridis} {et~al.}(2014){Koulouridis}, {Plionis}, {Melnyk},
  {Elyiv}, {Georgantopoulos}, {Clerc}, {Surdej}, {Chiappetti}, \&
  {Pierre}}]{koulouridis_2014}
{Koulouridis}, E., {Plionis}, M., {Melnyk}, O., {et~al.} 2014,
  \href{http://dx.doi.org/10.1051/0004-6361/201423601}{\JournalTitle{\aap},
  567, A83}

\bibitem[{{Laigle} {et~al.}(2018){Laigle}, {Pichon}, {Arnouts}, {McCracken},
  {Dubois}, {Devriendt}, {Slyz}, {Le Borgne}, {Benoit-L{\'e}vy}, {Hwang},
  {Ilbert}, {Kraljic}, {Malavasi}, {Park}, \& {Vibert}}]{laigle_2018}
{Laigle}, C., {Pichon}, C., {Arnouts}, S., {et~al.} 2018,
  \href{http://dx.doi.org/10.1093/mnras/stx3055}{\JournalTitle{\mnras}, 474,
  5437}

\bibitem[{{Lehmer} {et~al.}(2009){Lehmer}, {Alexander}, {Geach}, {Smail},
  {Basu-Zych}, {Bauer}, {Chapman}, {Matsuda}, {Scharf}, {Volonteri}, \&
  {Yamada}}]{lehmer_2009}
{Lehmer}, B.~D., {Alexander}, D.~M., {Geach}, J.~E., {et~al.} 2009,
  \href{http://dx.doi.org/10.1088/0004-637X/691/1/687}{\JournalTitle{\apj},
  691, 687}

\bibitem[{{Lehmer} {et~al.}(2012){Lehmer}, {Xue}, {Brandt}, {Alexander},
  {Bauer}, {Brusa}, {Comastri}, {Gilli}, {Hornschemeier}, {Luo}, {Paolillo},
  {Ptak}, {Shemmer}, {Schneider}, {Tozzi}, \& {Vignali}}]{lehmer_2012}
{Lehmer}, B.~D., {Xue}, Y.~Q., {Brandt}, W.~N., {et~al.} 2012,
  \href{http://dx.doi.org/10.1088/0004-637X/752/1/46}{\JournalTitle{\apj}, 752,
  46}

\bibitem[{{Lehmer} {et~al.}(2013){Lehmer}, {Lucy}, {Alexander}, {Best},
  {Geach}, {Harrison}, {Hornschemeier}, {Matsuda}, {Mullaney}, {Smail},
  {Sobral}, \& {Swinbank}}]{lehmer_2013}
{Lehmer}, B.~D., {Lucy}, A.~B., {Alexander}, D.~M., {et~al.} 2013,
  \href{http://dx.doi.org/10.1088/0004-637X/765/2/87}{\JournalTitle{\apj}, 765,
  87}

\bibitem[{{Li} {et~al.}(2019){Li}, {Xue}, {Sun}, {Liu}, {Vito}, {Brandt},
  {Hughes}, {Yang}, {Tozzi}, {Zhu}, {Zheng}, {Luo}, {Chen}, {Vignali}, {Gilli},
  \& {Shu}}]{li_2019}
{Li}, J., {Xue}, Y., {Sun}, M., {et~al.} 2019,
  \href{http://dx.doi.org/10.3847/1538-4357/ab184b}{\JournalTitle{\apj}, 877,
  5}

\bibitem[{{Lietzen} {et~al.}(2009){Lietzen}, {Hein{\"a}m{\"a}ki}, {Nurmi},
  {Tago}, {Saar}, {Liivam{\"a}gi}, {Tempel}, {Einasto}, {Einasto}, {Gramann},
  \& {Takalo}}]{lietzen_2009}
{Lietzen}, H., {Hein{\"a}m{\"a}ki}, P., {Nurmi}, P., {et~al.} 2009,
  \href{http://dx.doi.org/10.1051/0004-6361/200911628}{\JournalTitle{\aap},
  501, 145}

\bibitem[{{Lovisari} {et~al.}(2017){Lovisari}, {Forman}, {Jones}, {Ettori},
  {Andrade-Santos}, {Arnaud}, {D{\'e}mocl{\`e}s}, {Pratt}, {Randall}, \&
  {Kraft}}]{lovisari_2017}
{Lovisari}, L., {Forman}, W.~R., {Jones}, C., {et~al.} 2017,
  \href{http://dx.doi.org/10.3847/1538-4357/aa855f}{\JournalTitle{\apj}, 846,
  51}

\bibitem[{{Luo} {et~al.}(2017){Luo}, {Brandt}, {Xue}, {Lehmer}, {Alexander},
  {Bauer}, {Vito}, {Yang}, {Basu-Zych}, {Comastri}, {Gilli}, {Gu},
  {Hornschemeier}, {Koekemoer}, {Liu}, {Mainieri}, {Paolillo}, {Ranalli},
  {Rosati}, {Schneider}, {Shemmer}, {Smail}, {Sun}, {Tozzi}, {Vignali}, \&
  {Wang}}]{luo_2017}
{Luo}, B., {Brandt}, W.~N., {Xue}, Y.~Q., {et~al.} 2017,
  \href{http://dx.doi.org/10.3847/1538-4365/228/1/2}{\JournalTitle{\apjs}, 228,
  2}

\bibitem[{{Macuga} {et~al.}(2019){Macuga}, {Martini}, {Miller}, {Brodwin},
  {Hayashi}, {Kodama}, {Koyama}, {Overzier}, {Shimakawa}, {Tadaki}, \&
  {Tanaka}}]{macuga_2019}
{Macuga}, M., {Martini}, P., {Miller}, E.~D., {et~al.} 2019,
  \href{http://dx.doi.org/10.3847/1538-4357/ab0746}{\JournalTitle{\apj}, 874,
  54}

\bibitem[{{Mamon}(1992)}]{mamon_1992}
{Mamon}, G.~A. 1992,
  \href{http://dx.doi.org/10.1086/186656}{\JournalTitle{\apjl}, 401, L3}

\bibitem[{{Markevitch} \& {Vikhlinin}(2007)}]{markevitch_2007}
{Markevitch}, M., \& {Vikhlinin}, A. 2007,
  \href{http://dx.doi.org/10.1016/j.physrep.2007.01.001}{\JournalTitle{\physrep},
  443, 1}

\bibitem[{{Marshall} {et~al.}(2018){Marshall}, {Shabala}, {Krause}, {Pimbblet},
  {Croton}, \& {Owers}}]{marshal_2018}
{Marshall}, M.~A., {Shabala}, S.~S., {Krause}, M. G.~H., {et~al.} 2018,
  \href{http://dx.doi.org/10.1093/mnras/stx2996}{\JournalTitle{\mnras}, 474,
  3615}

\bibitem[{{Martini} {et~al.}(2009){Martini}, {Sivakoff}, \&
  {Mulchaey}}]{martini_2009}
{Martini}, P., {Sivakoff}, G.~R., \& {Mulchaey}, J.~S. 2009,
  \href{http://dx.doi.org/10.1088/0004-637X/701/1/66}{\JournalTitle{\apj}, 701,
  66}

\bibitem[{{Martini} {et~al.}(2013){Martini}, {Miller}, {Brodwin}, {Stanford},
  {Gonzalez}, {Bautz}, {Hickox}, {Stern}, {Eisenhardt}, {Galametz}, {Norman},
  {Jannuzi}, {Dey}, {Murray}, {Jones}, \& {Brown}}]{martini_2013}
{Martini}, P., {Miller}, E.~D., {Brodwin}, M., {et~al.} 2013,
  \href{http://dx.doi.org/10.1088/0004-637X/768/1/1}{\JournalTitle{\apj}, 768,
  1}

\bibitem[{{Miller} \& {Owen}(2003)}]{miller_2003}
{Miller}, N.~A., \& {Owen}, F.~N. 2003,
  \href{http://dx.doi.org/10.1086/374767}{\JournalTitle{\aj}, 125, 2427}

\bibitem[{{Mo} {et~al.}(2018){Mo}, {Gonzalez}, {Stern}, {Brodwin}, {Decker},
  {Eisenhardt}, {Moravec}, {Stanford}, \& {Wylezalek}}]{mo_2018}
{Mo}, W., {Gonzalez}, A., {Stern}, D., {et~al.} 2018,
  \href{http://dx.doi.org/10.3847/1538-4357/aaef83}{\JournalTitle{\apj}, 869,
  131}

\bibitem[{{Mohr} {et~al.}(1993){Mohr}, {Fabricant}, \& {Geller}}]{mohr_1993}
{Mohr}, J.~J., {Fabricant}, D.~G., \& {Geller}, M.~J. 1993,
  \href{http://dx.doi.org/10.1086/173019}{\JournalTitle{\apj}, 413, 492}

\bibitem[{{Moore} {et~al.}(1996){Moore}, {Katz}, {Lake}, {Dressler}, \&
  {Oemler}}]{moore_1996}
{Moore}, B., {Katz}, N., {Lake}, G., {Dressler}, A., \& {Oemler}, A. 1996,
  \href{http://dx.doi.org/10.1038/379613a0}{\JournalTitle{\nat}, 379, 613}

\bibitem[{{Moravec} {et~al.}(2020){Moravec}, {Gonzalez}, {Dicker}, {Alberts},
  {Brodwin}, {Clarke}, {Connor}, {Decker}, {Devlin}, {Eisenhardt}, {Mason},
  {Mo}, {Mroczkowski}, {Pope}, {Romero}, {Sarazin}, {Sievers}, {Stanford},
  {Stern}, {Wylezalek}, \& {Zago}}]{moravec_2020}
{Moravec}, E., {Gonzalez}, A.~H., {Dicker}, S., {et~al.} 2020,
  \href{http://dx.doi.org/10.3847/1538-4357/aba0b2}{\JournalTitle{\apj}, 898,
  145}

\bibitem[{{Moster} {et~al.}(2011){Moster}, {Somerville}, {Newman}, \&
  {Rix}}]{moster_2011}
{Moster}, B.~P., {Somerville}, R.~S., {Newman}, J.~A., \& {Rix}, H.-W. 2011,
  \href{http://dx.doi.org/10.1088/0004-637X/731/2/113}{\JournalTitle{\apj},
  731, 113}

\bibitem[{{Mu{\~n}oz Rodr{\'\i}guez} {et~al.}(2023){Mu{\~n}oz Rodr{\'\i}guez},
  {Georgakakis}, {Shankar}, {Allevato}, {Bonoli}, {Brusa}, {Lapi}, \&
  {Viitanen}}]{munoz_2023}
{Mu{\~n}oz Rodr{\'\i}guez}, I., {Georgakakis}, A., {Shankar}, F., {et~al.}
  2023, \href{http://dx.doi.org/10.1093/mnras/stac3114}{\JournalTitle{\mnras},
  518, 1041}

\bibitem[{{Noordeh} {et~al.}(2020){Noordeh}, {Canning}, {King}, {Allen},
  {Mantz}, {Morris}, {Ehlert}, {von der Linden}, {Brandt}, {Luo}, {Xue}, \&
  {Kelly}}]{noordeh_2020}
{Noordeh}, E., {Canning}, R.~E.~A., {King}, A., {et~al.} 2020,
  \href{http://dx.doi.org/10.1093/mnras/staa2682}{\JournalTitle{\mnras}, 498,
  4095}

\bibitem[{{Nurgaliev} {et~al.}(2013){Nurgaliev}, {McDonald}, {Benson},
  {Miller}, {Stubbs}, \& {Vikhlinin}}]{nurgalive_2013}
{Nurgaliev}, D., {McDonald}, M., {Benson}, B.~A., {et~al.} 2013,
  \href{http://dx.doi.org/10.1088/0004-637X/779/2/112}{\JournalTitle{\apj},
  779, 112}

\bibitem[{{Nurgaliev} {et~al.}(2017){Nurgaliev}, {McDonald}, {Benson}, {Bleem},
  {Bocquet}, {Forman}, {Garmire}, {Gupta}, {Hlavacek-Larrondo}, {Mohr},
  {Nagai}, {Rapetti}, {Stark}, {Stubbs}, \& {Vikhlinin}}]{nurgaliev_2017}
{Nurgaliev}, D., {McDonald}, M., {Benson}, B.~A., {et~al.} 2017,
  \href{http://dx.doi.org/10.3847/1538-4357/aa6db4}{\JournalTitle{\apj}, 841,
  5}

\bibitem[{{Overzier}(2016)}]{Overzier_2016}
{Overzier}, R.~A. 2016,
  \href{http://dx.doi.org/10.1007/s00159-016-0100-3}{\JournalTitle{\aapr}, 24,
  14}

\bibitem[{{Owen} {et~al.}(1999){Owen}, {Ledlow}, {Keel}, \&
  {Morrison}}]{owen_1999}
{Owen}, F.~N., {Ledlow}, M.~J., {Keel}, W.~C., \& {Morrison}, G.~E. 1999,
  \href{http://dx.doi.org/10.1086/300974}{\JournalTitle{\aj}, 118, 633}

\bibitem[{{Padovani}(2016)}]{padovani_2016}
{Padovani}, P. 2016,
  \href{http://dx.doi.org/10.1007/s00159-016-0098-6}{\JournalTitle{\aapr}, 24,
  13}

\bibitem[{{Parekh} {et~al.}(2015){Parekh}, {van der Heyden}, {Ferrari},
  {Angus}, \& {Holwerda}}]{parekh_2015}
{Parekh}, V., {van der Heyden}, K., {Ferrari}, C., {Angus}, G., \& {Holwerda},
  B. 2015,
  \href{http://dx.doi.org/10.1051/0004-6361/201424123}{\JournalTitle{\aap},
  575, A127}

\bibitem[{{Pimbblet} {et~al.}(2013){Pimbblet}, {Shabala}, {Haines},
  {Fraser-McKelvie}, \& {Floyd}}]{Pimbblet_2013}
{Pimbblet}, K.~A., {Shabala}, S.~S., {Haines}, C.~P., {Fraser-McKelvie}, A., \&
  {Floyd}, D.~J.~E. 2013,
  \href{http://dx.doi.org/10.1093/mnras/sts470}{\JournalTitle{\mnras}, 429,
  1827}

\bibitem[{{Planck Collaboration} {et~al.}(2016{\natexlab{a}}){Planck
  Collaboration}, {Ade}, {Aghanim}, {Arnaud}, {Ashdown}, {Aumont},
  {Baccigalupi}, {Banday}, {Barreiro}, {Bartlett}, {Bartolo}, {Battaner},
  {Battye}, {Benabed}, {Beno{\^\i}t}, {Benoit-L{\'e}vy}, {Bernard},
  {Bersanelli}, {Bielewicz}, {Bock}, {Bonaldi}, {Bonavera}, {Bond}, {Borrill},
  {Bouchet}, {Boulanger}, {Bucher}, {Burigana}, {Butler}, {Calabrese},
  {Cardoso}, {Catalano}, {Challinor}, {Chamballu}, {Chary}, {Chiang}, {Chluba},
  {Christensen}, {Church}, {Clements}, {Colombi}, {Colombo}, {Combet},
  {Coulais}, {Crill}, {Curto}, {Cuttaia}, {Danese}, {Davies}, {Davis}, {de
  Bernardis}, {de Rosa}, {de Zotti}, {Delabrouille}, {D{\'e}sert}, {Di
  Valentino}, {Dickinson}, {Diego}, {Dolag}, {Dole}, {Donzelli}, {Dor{\'e}},
  {Douspis}, {Ducout}, {Dunkley}, {Dupac}, {Efstathiou}, {Elsner},
  {En{\ss}lin}, {Eriksen}, {Farhang}, {Fergusson}, {Finelli}, {Forni},
  {Frailis}, {Fraisse}, {Franceschi}, {Frejsel}, {Galeotta}, {Galli}, {Ganga},
  {Gauthier}, {Gerbino}, {Ghosh}, {Giard}, {Giraud-H{\'e}raud}, {Giusarma},
  {Gjerl{\o}w}, {Gonz{\'a}lez-Nuevo}, {G{\'o}rski}, {Gratton}, {Gregorio},
  {Gruppuso}, {Gudmundsson}, {Hamann}, {Hansen}, {Hanson}, {Harrison}, {Helou},
  {Henrot-Versill{\'e}}, {Hern{\'a}ndez-Monteagudo}, {Herranz}, {Hildebrandt},
  {Hivon}, {Hobson}, {Holmes}, {Hornstrup}, {Hovest}, {Huang}, {Huffenberger},
  {Hurier}, {Jaffe}, {Jaffe}, {Jones}, {Juvela}, {Keih{\"a}nen}, {Keskitalo},
  {Kisner}, {Kneissl}, {Knoche}, {Knox}, {Kunz}, {Kurki-Suonio}, {Lagache},
  {L{\"a}hteenm{\"a}ki}, {Lamarre}, {Lasenby}, {Lattanzi}, {Lawrence}, {Leahy},
  {Leonardi}, {Lesgourgues}, {Levrier}, {Lewis}, {Liguori}, {Lilje},
  {Linden-V{\o}rnle}, {L{\'o}pez-Caniego}, {Lubin}, {Mac{\'\i}as-P{\'e}rez},
  {Maggio}, {Maino}, {Mandolesi}, {Mangilli}, {Marchini}, {Maris}, {Martin},
  {Martinelli}, {Mart{\'\i}nez-Gonz{\'a}lez}, {Masi}, {Matarrese}, {McGehee},
  {Meinhold}, {Melchiorri}, {Melin}, {Mendes}, {Mennella}, {Migliaccio},
  {Millea}, {Mitra}, {Miville-Desch{\^e}nes}, {Moneti}, {Montier}, {Morgante},
  {Mortlock}, {Moss}, {Munshi}, {Murphy}, {Naselsky}, {Nati}, {Natoli},
  {Netterfield}, {N{\o}rgaard-Nielsen}, {Noviello}, {Novikov}, {Novikov},
  {Oxborrow}, {Paci}, {Pagano}, {Pajot}, {Paladini}, {Paoletti}, {Partridge},
  {Pasian}, {Patanchon}, {Pearson}, {Perdereau}, {Perotto}, {Perrotta},
  {Pettorino}, {Piacentini}, {Piat}, {Pierpaoli}, {Pietrobon}, {Plaszczynski},
  {Pointecouteau}, {Polenta}, {Popa}, {Pratt}, {Pr{\'e}zeau}, {Prunet},
  {Puget}, {Rachen}, {Reach}, {Rebolo}, {Reinecke}, {Remazeilles}, {Renault},
  {Renzi}, {Ristorcelli}, {Rocha}, {Rosset}, {Rossetti}, {Roudier},
  {Rouill{\'e} d'Orfeuil}, {Rowan-Robinson}, {Rubi{\~n}o-Mart{\'\i}n},
  {Rusholme}, {Said}, {Salvatelli}, {Salvati}, {Sandri}, {Santos},
  {Savelainen}, {Savini}, {Scott}, {Seiffert}, {Serra}, {Shellard}, {Spencer},
  {Spinelli}, {Stolyarov}, {Stompor}, {Sudiwala}, {Sunyaev}, {Sutton},
  {Suur-Uski}, {Sygnet}, {Tauber}, {Terenzi}, {Toffolatti}, {Tomasi},
  {Tristram}, {Trombetti}, {Tucci}, {Tuovinen}, {T{\"u}rler}, {Umana},
  {Valenziano}, {Valiviita}, {Van Tent}, {Vielva}, {Villa}, {Wade}, {Wandelt},
  {Wehus}, {White}, {White}, {Wilkinson}, {Yvon}, {Zacchei}, \&
  {Zonca}}]{planck_2016}
{Planck Collaboration}, {Ade}, P.~A.~R., {Aghanim}, N., {et~al.}
  2016{\natexlab{a}},
  \href{http://dx.doi.org/10.1051/0004-6361/201525830}{\JournalTitle{\aap},
  594, A13}

\bibitem[{{Planck Collaboration} {et~al.}(2016{\natexlab{b}}){Planck
  Collaboration}, {Ade}, {Aghanim}, {Arnaud}, {Ashdown}, {Aumont},
  {Baccigalupi}, {Banday}, {Barreiro}, {Barrena}, {Bartlett}, {Bartolo},
  {Battaner}, {Battye}, {Benabed}, {Beno{\^\i}t}, {Benoit-L{\'e}vy}, {Bernard},
  {Bersanelli}, {Bielewicz}, {Bikmaev}, {B{\"o}hringer}, {Bonaldi}, {Bonavera},
  {Bond}, {Borrill}, {Bouchet}, {Bucher}, {Burenin}, {Burigana}, {Butler},
  {Calabrese}, {Cardoso}, {Carvalho}, {Catalano}, {Challinor}, {Chamballu},
  {Chary}, {Chiang}, {Chon}, {Christensen}, {Clements}, {Colombi}, {Colombo},
  {Combet}, {Comis}, {Couchot}, {Coulais}, {Crill}, {Curto}, {Cuttaia},
  {Dahle}, {Danese}, {Davies}, {Davis}, {de Bernardis}, {de Rosa}, {de Zotti},
  {Delabrouille}, {D{\'e}sert}, {Dickinson}, {Diego}, {Dolag}, {Dole},
  {Donzelli}, {Dor{\'e}}, {Douspis}, {Ducout}, {Dupac}, {Efstathiou},
  {Eisenhardt}, {Elsner}, {En{\ss}lin}, {Eriksen}, {Falgarone}, {Fergusson},
  {Feroz}, {Ferragamo}, {Finelli}, {Forni}, {Frailis}, {Fraisse}, {Franceschi},
  {Frejsel}, {Galeotta}, {Galli}, {Ganga}, {G{\'e}nova-Santos}, {Giard},
  {Giraud-H{\'e}raud}, {Gjerl{\o}w}, {Gonz{\'a}lez-Nuevo}, {G{\'o}rski},
  {Grainge}, {Gratton}, {Gregorio}, {Gruppuso}, {Gudmundsson}, {Hansen},
  {Hanson}, {Harrison}, {Hempel}, {Henrot-Versill{\'e}},
  {Hern{\'a}ndez-Monteagudo}, {Herranz}, {Hildebrandt}, {Hivon}, {Hobson},
  {Holmes}, {Hornstrup}, {Hovest}, {Huffenberger}, {Hurier}, {Jaffe}, {Jaffe},
  {Jin}, {Jones}, {Juvela}, {Keih{\"a}nen}, {Keskitalo}, {Khamitov}, {Kisner},
  {Kneissl}, {Knoche}, {Kunz}, {Kurki-Suonio}, {Lagache}, {Lamarre}, {Lasenby},
  {Lattanzi}, {Lawrence}, {Leonardi}, {Lesgourgues}, {Levrier}, {Liguori},
  {Lilje}, {Linden-V{\o}rnle}, {L{\'o}pez-Caniego}, {Lubin},
  {Mac{\'\i}as-P{\'e}rez}, {Maggio}, {Maino}, {Mak}, {Mandolesi}, {Mangilli},
  {Martin}, {Mart{\'\i}nez-Gonz{\'a}lez}, {Masi}, {Matarrese}, {Mazzotta},
  {McGehee}, {Mei}, {Melchiorri}, {Melin}, {Mendes}, {Mennella}, {Migliaccio},
  {Mitra}, {Miville-Desch{\^e}nes}, {Moneti}, {Montier}, {Morgante},
  {Mortlock}, {Moss}, {Munshi}, {Murphy}, {Naselsky}, {Nastasi}, {Nati},
  {Natoli}, {Netterfield}, {N{\o}rgaard-Nielsen}, {Noviello}, {Novikov},
  {Novikov}, {Olamaie}, {Oxborrow}, {Paci}, {Pagano}, {Pajot}, {Paoletti},
  {Pasian}, {Patanchon}, {Pearson}, {Perdereau}, {Perotto}, {Perrott},
  {Perrotta}, {Pettorino}, {Piacentini}, {Piat}, {Pierpaoli}, {Pietrobon},
  {Plaszczynski}, {Pointecouteau}, {Polenta}, {Pratt}, {Pr{\'e}zeau}, {Prunet},
  {Puget}, {Rachen}, {Reach}, {Rebolo}, {Reinecke}, {Remazeilles}, {Renault},
  {Renzi}, {Ristorcelli}, {Rocha}, {Rosset}, {Rossetti}, {Roudier}, {Rozo},
  {Rubi{\~n}o-Mart{\'\i}n}, {Rumsey}, {Rusholme}, {Rykoff}, {Sandri}, {Santos},
  {Saunders}, {Savelainen}, {Savini}, {Schammel}, {Scott}, {Seiffert},
  {Shellard}, {Shimwell}, {Spencer}, {Stanford}, {Stern}, {Stolyarov},
  {Stompor}, {Streblyanska}, {Sudiwala}, {Sunyaev}, {Sutton}, {Suur-Uski},
  {Sygnet}, {Tauber}, {Terenzi}, {Toffolatti}, {Tomasi}, {Tramonte},
  {Tristram}, {Tucci}, {Tuovinen}, {Umana}, {Valenziano}, {Valiviita}, {Van
  Tent}, {Vielva}, {Villa}, {Wade}, {Wandelt}, {Wehus}, {White}, {Wright},
  {Yvon}, {Zacchei}, \& {Zonca}}]{planck_2015}
{Planck Collaboration}, {Ade}, P.~A.~R., {Aghanim}, N., {et~al.}
  2016{\natexlab{b}},
  \href{http://dx.doi.org/10.1051/0004-6361/201525823}{\JournalTitle{\aap},
  594, A27}

\bibitem[{{Poggianti} {et~al.}(2017{\natexlab{a}}){Poggianti}, {Moretti},
  {Gullieuszik}, {Fritz}, {Jaff{\'e}}, {Bettoni}, {Fasano}, {Bellhouse}, {Hau},
  {Vulcani}, {Biviano}, {Omizzolo}, {Paccagnella}, {D'Onofrio}, {Cava},
  {Sheen}, {Couch}, \& {Owers}}]{poggianti_2017}
{Poggianti}, B.~M., {Moretti}, A., {Gullieuszik}, M., {et~al.}
  2017{\natexlab{a}},
  \href{http://dx.doi.org/10.3847/1538-4357/aa78ed}{\JournalTitle{\apj}, 844,
  48}

\bibitem[{{Poggianti} {et~al.}(2017{\natexlab{b}}){Poggianti}, {Jaff{\'e}},
  {Moretti}, {Gullieuszik}, {Radovich}, {Tonnesen}, {Fritz}, {Bettoni},
  {Vulcani}, {Fasano}, {Bellhouse}, {Hau}, \& {Omizzolo}}]{poggianti_2017_b}
{Poggianti}, B.~M., {Jaff{\'e}}, Y.~L., {Moretti}, A., {et~al.}
  2017{\natexlab{b}},
  \href{http://dx.doi.org/10.1038/nature23462}{\JournalTitle{\nat}, 548, 304}

\bibitem[{{Poole} {et~al.}(2007){Poole}, {Babul}, {McCarthy}, {Fardal},
  {Bildfell}, {Quinn}, \& {Mahdavi}}]{poole_2007}
{Poole}, G.~B., {Babul}, A., {McCarthy}, I.~G., {et~al.} 2007,
  \href{http://dx.doi.org/10.1111/j.1365-2966.2007.12107.x}{\JournalTitle{\mnras},
  380, 437}

\bibitem[{{Poole} {et~al.}(2006){Poole}, {Fardal}, {Babul}, {McCarthy},
  {Quinn}, \& {Wadsley}}]{poole_2006}
{Poole}, G.~B., {Fardal}, M.~A., {Babul}, A., {et~al.} 2006,
  \href{http://dx.doi.org/10.1111/j.1365-2966.2006.10916.x}{\JournalTitle{\mnras},
  373, 881}

\bibitem[{{Rasia} {et~al.}(2013){Rasia}, {Meneghetti}, \&
  {Ettori}}]{raisa_2013}
{Rasia}, E., {Meneghetti}, M., \& {Ettori}, S. 2013,
  \href{http://dx.doi.org/10.1080/21672857.2013.11519713}{\JournalTitle{The
  Astronomical Review}, 8, 40}

\bibitem[{{Ricarte} {et~al.}(2020){Ricarte}, {Tremmel}, {Natarajan}, \&
  {Quinn}}]{ricarte_2020}
{Ricarte}, A., {Tremmel}, M., {Natarajan}, P., \& {Quinn}, T. 2020,
  \href{http://dx.doi.org/10.3847/2041-8213/ab9022}{\JournalTitle{\apjl}, 895,
  L8}

\bibitem[{{Ruderman} \& {Ebeling}(2005)}]{ruderman_2005}
{Ruderman}, J.~T., \& {Ebeling}, H. 2005,
  \href{http://dx.doi.org/10.1086/430131}{\JournalTitle{\apjl}, 623, L81}

\bibitem[{{Ruppin} {et~al.}(2020){Ruppin}, {McDonald}, {Brodwin}, {Adam},
  {Ade}, {Andr{\'e}}, {Andrianasolo}, {Arnaud}, {Aussel}, {Bartalucci},
  {Bautz}, {Beelen}, {Beno{\^\i}t}, {Bideaud}, {Bourrion}, {Calvo}, {Catalano},
  {Comis}, {Decker}, {De Petris}, {D{\'e}sert}, {Doyle}, {Driessen},
  {Eisenhardt}, {Gomez}, {Gonzalez}, {Goupy}, {K{\'e}ruzor{\'e}}, {Kramer},
  {Ladjelate}, {Lagache}, {Leclercq}, {Lestrade}, {Mac{\'\i}as-P{\'e}rez},
  {Mauskopf}, {Mayet}, {Monfardini}, {Moravec}, {Perotto}, {Pisano},
  {Pointecouteau}, {Ponthieu}, {Pratt}, {Rev{\'e}ret}, {Ritacco}, {Romero},
  {Roussel}, {Schuster}, {Shu}, {Sievers}, {Stanford}, {Stern}, {Tucker}, \&
  {Zylka}}]{ruppin_2020}
{Ruppin}, F., {McDonald}, M., {Brodwin}, M., {et~al.} 2020,
  \href{http://dx.doi.org/10.3847/1538-4357/ab8007}{\JournalTitle{\apj}, 893,
  74}

\bibitem[{{Rykoff} {et~al.}(2016){Rykoff}, {Rozo}, {Hollowood},
  {Bermeo-Hernandez}, {Jeltema}, {Mayers}, {Romer}, {Rooney}, {Saro}, {Vergara
  Cervantes}, {Wechsler}, {Wilcox}, {Abbott}, {Abdalla}, {Allam}, {Annis},
  {Benoit-L{\'e}vy}, {Bernstein}, {Bertin}, {Brooks}, {Burke}, {Capozzi},
  {Carnero Rosell}, {Carrasco Kind}, {Castander}, {Childress}, {Collins},
  {Cunha}, {D'Andrea}, {da Costa}, {Davis}, {Desai}, {Diehl}, {Dietrich},
  {Doel}, {Evrard}, {Finley}, {Flaugher}, {Fosalba}, {Frieman}, {Glazebrook},
  {Goldstein}, {Gruen}, {Gruendl}, {Gutierrez}, {Hilton}, {Honscheid}, {Hoyle},
  {James}, {Kay}, {Kuehn}, {Kuropatkin}, {Lahav}, {Lewis}, {Lidman}, {Lima},
  {Maia}, {Mann}, {Marshall}, {Martini}, {Melchior}, {Miller}, {Miquel},
  {Mohr}, {Nichol}, {Nord}, {Ogando}, {Plazas}, {Reil}, {Sahl{\'e}n},
  {Sanchez}, {Santiago}, {Scarpine}, {Schubnell}, {Sevilla-Noarbe}, {Smith},
  {Soares-Santos}, {Sobreira}, {Stott}, {Suchyta}, {Swanson}, {Tarle},
  {Thomas}, {Tucker}, {Uddin}, {Viana}, {Vikram}, {Walker}, {Zhang}, \& {DES
  Collaboration}}]{Rykoff_2016}
{Rykoff}, E.~S., {Rozo}, E., {Hollowood}, D., {et~al.} 2016,
  \href{http://dx.doi.org/10.3847/0067-0049/224/1/1}{\JournalTitle{\apjs}, 224,
  1}

\bibitem[{{Santini} {et~al.}(2014){Santini}, {Maiolino}, {Magnelli}, {Lutz},
  {Lamastra}, {Li Causi}, {Eales}, {Andreani}, {Berta}, {Buat}, {Cooray},
  {Cresci}, {Daddi}, {Farrah}, {Fontana}, {Franceschini}, {Genzel}, {Granato},
  {Grazian}, {Le Floc'h}, {Magdis}, {Magliocchetti}, {Mannucci}, {Menci},
  {Nordon}, {Oliver}, {Popesso}, {Pozzi}, {Riguccini}, {Rodighiero}, {Rosario},
  {Salvato}, {Scott}, {Silva}, {Tacconi}, {Viero}, {Wang}, {Wuyts}, \&
  {Xu}}]{santini_2014}
{Santini}, P., {Maiolino}, R., {Magnelli}, B., {et~al.} 2014,
  \href{http://dx.doi.org/10.1051/0004-6361/201322835}{\JournalTitle{\aap},
  562, A30}

\bibitem[{{Schulz} \& {Struck}(2001)}]{schulz_2001}
{Schulz}, S., \& {Struck}, C. 2001,
  \href{http://dx.doi.org/10.1046/j.1365-8711.2001.04847.x}{\JournalTitle{\mnras},
  328, 185}

\bibitem[{{Scoville} {et~al.}(2013){Scoville}, {Arnouts}, {Aussel}, {Benson},
  {Bongiorno}, {Bundy}, {Calvo}, {Capak}, {Carollo}, {Civano}, {Dunlop},
  {Elvis}, {Faisst}, {Finoguenov}, {Fu}, {Giavalisco}, {Guo}, {Ilbert},
  {Iovino}, {Kajisawa}, {Kartaltepe}, {Leauthaud}, {Le F{\`e}vre}, {LeFloch},
  {Lilly}, {Liu}, {Manohar}, {Massey}, {Masters}, {McCracken}, {Mobasher},
  {Peng}, {Renzini}, {Rhodes}, {Salvato}, {Sanders}, {Sarvestani}, {Scarlata},
  {Schinnerer}, {Sheth}, {Shopbell}, {Smol{\v{c}}i{\'c}}, {Taniguchi},
  {Taylor}, {White}, \& {Yan}}]{scoville_2013}
{Scoville}, N., {Arnouts}, S., {Aussel}, H., {et~al.} 2013,
  \href{http://dx.doi.org/10.1088/0067-0049/206/1/3}{\JournalTitle{\apjs}, 206,
  3}

\bibitem[{{Silverman} {et~al.}(2008){Silverman}, {Mainieri}, {Lehmer},
  {Alexander}, {Bauer}, {Bergeron}, {Brandt}, {Gilli}, {Hasinger}, {Schneider},
  {Tozzi}, {Vignali}, {Koekemoer}, {Miyaji}, {Popesso}, {Rosati}, \&
  {Szokoly}}]{silverman_2008}
{Silverman}, J.~D., {Mainieri}, V., {Lehmer}, B.~D., {et~al.} 2008,
  \href{http://dx.doi.org/10.1086/527283}{\JournalTitle{\apj}, 675, 1025}

\bibitem[{{Sobral} {et~al.}(2015){Sobral}, {Stroe}, {Dawson}, {Wittman}, {Jee},
  {R{\"o}ttgering}, {van Weeren}, \& {Br{\"u}ggen}}]{sobral_2015}
{Sobral}, D., {Stroe}, A., {Dawson}, W.~A., {et~al.} 2015,
  \href{http://dx.doi.org/10.1093/mnras/stv521}{\JournalTitle{\mnras}, 450,
  630}

\bibitem[{{Somerville} {et~al.}(2004){Somerville}, {Lee}, {Ferguson},
  {Gardner}, {Moustakas}, \& {Giavalisco}}]{somerville_2004}
{Somerville}, R.~S., {Lee}, K., {Ferguson}, H.~C., {et~al.} 2004,
  \href{http://dx.doi.org/10.1086/378628}{\JournalTitle{\apjl}, 600, L171}

\bibitem[{{Stanford} {et~al.}(2014){Stanford}, {Gonzalez}, {Brodwin},
  {Gettings}, {Eisenhardt}, {Stern}, \& {Wylezalek}}]{stanford_2014}
{Stanford}, S.~A., {Gonzalez}, A.~H., {Brodwin}, M., {et~al.} 2014,
  \href{http://dx.doi.org/10.1088/0067-0049/213/2/25}{\JournalTitle{\apjs},
  213, 25}

\bibitem[{{Strand} {et~al.}(2008){Strand}, {Brunner}, \& {Myers}}]{strand_2008}
{Strand}, N.~E., {Brunner}, R.~J., \& {Myers}, A.~D. 2008,
  \href{http://dx.doi.org/10.1086/592099}{\JournalTitle{\apj}, 688, 180}

\bibitem[{{Stroe} \& {Sobral}(2021)}]{stroe_2021}
{Stroe}, A., \& {Sobral}, D. 2021,
  \href{http://dx.doi.org/10.3847/1538-4357/abe7f8}{\JournalTitle{\apj}, 912,
  55}

\bibitem[{{Stroe} {et~al.}(2017){Stroe}, {Sobral}, {Paulino-Afonso}, {Alegre},
  {Calhau}, {Santos}, \& {van Weeren}}]{stroe_2017}
{Stroe}, A., {Sobral}, D., {Paulino-Afonso}, A., {et~al.} 2017,
  \href{http://dx.doi.org/10.1093/mnras/stw2939}{\JournalTitle{\mnras}, 465,
  2916}

\bibitem[{{Stroe} {et~al.}(2015){Stroe}, {Sobral}, {Dawson}, {Jee}, {Hoekstra},
  {Wittman}, {van Weeren}, {Br{\"u}ggen}, \& {R{\"o}ttgering}}]{stroe_2015}
{Stroe}, A., {Sobral}, D., {Dawson}, W., {et~al.} 2015,
  \href{http://dx.doi.org/10.1093/mnras/stu2519}{\JournalTitle{\mnras}, 450,
  646}

\bibitem[{{Tacconi} {et~al.}(2010){Tacconi}, {Genzel}, {Neri}, {Cox}, {Cooper},
  {Shapiro}, {Bolatto}, {Bouch{\'e}}, {Bournaud}, {Burkert}, {Combes},
  {Comerford}, {Davis}, {F{\"o}rster Schreiber}, {Garcia-Burillo},
  {Gracia-Carpio}, {Lutz}, {Naab}, {Omont}, {Shapley}, {Sternberg}, \&
  {Weiner}}]{tacconi_2010}
{Tacconi}, L.~J., {Genzel}, R., {Neri}, R., {et~al.} 2010,
  \href{http://dx.doi.org/10.1038/nature08773}{\JournalTitle{\nat}, 463, 781}

\bibitem[{{Tonnesen} \& {Bryan}(2009)}]{tonnesen_2009}
{Tonnesen}, S., \& {Bryan}, G.~L. 2009,
  \href{http://dx.doi.org/10.1088/0004-637X/694/2/789}{\JournalTitle{\apj},
  694, 789}

\bibitem[{{Umehata} {et~al.}(2014){Umehata}, {Tamura}, {Kohno}, {Hatsukade},
  {Scott}, {Kubo}, {Yamada}, {Ivison}, {Cybulski}, {Aretxaga}, {Austermann},
  {Hughes}, {Ezawa}, {Hayashino}, {Ikarashi}, {Iono}, {Kawabe}, {Matsuda},
  {Matsuo}, {Nakanishi}, {Oshima}, {Perera}, {Takata}, {Wilson}, \&
  {Yun}}]{umehata_2014}
{Umehata}, H., {Tamura}, Y., {Kohno}, K., {et~al.} 2014,
  \href{http://dx.doi.org/10.1093/mnras/stu447}{\JournalTitle{\mnras}, 440,
  3462}

\bibitem[{{van Weeren} {et~al.}(2019){van Weeren}, {de Gasperin}, {Akamatsu},
  {Br{\"u}ggen}, {Feretti}, {Kang}, {Stroe}, \& {Zandanel}}]{weeren_2019}
{van Weeren}, R.~J., {de Gasperin}, F., {Akamatsu}, H., {et~al.} 2019,
  \href{http://dx.doi.org/10.1007/s11214-019-0584-z}{\JournalTitle{\ssr}, 215,
  16}

\bibitem[{{Vikhlinin} {et~al.}(1998){Vikhlinin}, {McNamara}, {Forman}, {Jones},
  {Quintana}, \& {Hornstrup}}]{Vikhlinin_1998}
{Vikhlinin}, A., {McNamara}, B.~R., {Forman}, W., {et~al.} 1998,
  \href{http://dx.doi.org/10.1086/311305}{\JournalTitle{\apjl}, 498, L21}

\bibitem[{{Vikhlinin} {et~al.}(2009){Vikhlinin}, {Burenin}, {Ebeling},
  {Forman}, {Hornstrup}, {Jones}, {Kravtsov}, {Murray}, {Nagai}, {Quintana}, \&
  {Voevodkin}}]{vikhlinin_2009}
{Vikhlinin}, A., {Burenin}, R.~A., {Ebeling}, H., {et~al.} 2009,
  \href{http://dx.doi.org/10.1088/0004-637X/692/2/1033}{\JournalTitle{\apj},
  692, 1033}

\bibitem[{Virtanen {et~al.}(2020)Virtanen, Gommers, Oliphant, Haberland, Reddy,
  Cournapeau, Burovski, Peterson, Weckesser, Bright, {van der Walt}, Brett,
  Wilson, Millman, Mayorov, Nelson, Jones, Kern, Larson, Carey, Polat, Feng,
  Moore, {VanderPlas}, Laxalde, Perktold, Cimrman, Henriksen, Quintero, Harris,
  Archibald, Ribeiro, Pedregosa, {van Mulbregt}, \& {SciPy 1.0
  Contributors}}]{scipy}
Virtanen, P., Gommers, R., Oliphant, T.~E., {et~al.} 2020,
  \href{http://dx.doi.org/10.1038/s41592-019-0686-2}{\JournalTitle{Nature
  Methods}, 17, 261}

\bibitem[{{von der Linden} {et~al.}(2014){von der Linden}, {Allen},
  {Applegate}, {Kelly}, {Allen}, {Ebeling}, {Burchat}, {Burke}, {Donovan},
  {Morris}, {Blandford}, {Erben}, \& {Mantz}}]{von_linden_2014}
{von der Linden}, A., {Allen}, M.~T., {Applegate}, D.~E., {et~al.} 2014,
  \href{http://dx.doi.org/10.1093/mnras/stt1945}{\JournalTitle{\mnras}, 439, 2}

\bibitem[{Waskom(2021)}]{seaborn}
Waskom, M.~L. 2021,
  \href{http://dx.doi.org/10.21105/joss.03021}{\JournalTitle{Journal of Open
  Source Software}, 6, 3021}

\bibitem[{Watson(1961)}]{watson_1961}
Watson, G.~S. 1961, \JournalTitle{Biometrika}, 48, 109

\bibitem[{{Wei{\ss}mann} {et~al.}(2013){Wei{\ss}mann}, {B{\"o}hringer}, \&
  {Chon}}]{weibmann_2013b}
{Wei{\ss}mann}, A., {B{\"o}hringer}, H., \& {Chon}, G. 2013,
  \href{http://dx.doi.org/10.1051/0004-6361/201321495}{\JournalTitle{\aap},
  555, A147}

\bibitem[{{W}es {M}c{K}inney(2010)}]{pandas}
{W}es {M}c{K}inney. 2010,
  \href{http://dx.doi.org/10.25080/Majora-92bf1922-00a}{in {P}roceedings of the
  9th {P}ython in {S}cience {C}onference, ed. {S}t\'efan van~der {W}alt \&
  {J}arrod {M}illman}, 56

\bibitem[{{Wright} {et~al.}(2010){Wright}, {Eisenhardt}, {Mainzer}, {Ressler},
  {Cutri}, {Jarrett}, {Kirkpatrick}, {Padgett}, {McMillan}, {Skrutskie},
  {Stanford}, {Cohen}, {Walker}, {Mather}, {Leisawitz}, {Gautier}, {McLean},
  {Benford}, {Lonsdale}, {Blain}, {Mendez}, {Irace}, {Duval}, {Liu}, {Royer},
  {Heinrichsen}, {Howard}, {Shannon}, {Kendall}, {Walsh}, {Larsen}, {Cardon},
  {Schick}, {Schwalm}, {Abid}, {Fabinsky}, {Naes}, \& {Tsai}}]{wright_2010}
{Wright}, E.~L., {Eisenhardt}, P. R.~M., {Mainzer}, A.~K., {et~al.} 2010,
  \href{http://dx.doi.org/10.1088/0004-6256/140/6/1868}{\JournalTitle{\aj},
  140, 1868}

\bibitem[{{Xue} {et~al.}(2011){Xue}, {Luo}, {Brandt}, {Bauer}, {Lehmer},
  {Broos}, {Schneider}, {Alexander}, {Brusa}, {Comastri}, {Fabian}, {Gilli},
  {Hasinger}, {Hornschemeier}, {Koekemoer}, {Liu}, {Mainieri}, {Paolillo},
  {Rafferty}, {Rosati}, {Shemmer}, {Silverman}, {Smail}, {Tozzi}, \&
  {Vignali}}]{xue_2011}
{Xue}, Y.~Q., {Luo}, B., {Brandt}, W.~N., {et~al.} 2011,
  \href{http://dx.doi.org/10.1088/0067-0049/195/1/10}{\JournalTitle{\apjs},
  195, 10}

\bibitem[{{Yang} {et~al.}(2018){Yang}, {Brandt}, {Darvish}, {Chen}, {Vito},
  {Alexander}, {Bauer}, \& {Trump}}]{yang_2018}
{Yang}, G., {Brandt}, W.~N., {Darvish}, B., {et~al.} 2018,
  \href{http://dx.doi.org/10.1093/mnras/sty1910}{\JournalTitle{\mnras}, 480,
  1022}

\bibitem[{{Yang} {et~al.}(2017){Yang}, {Chen}, {Vito}, {Brandt}, {Alexander},
  {Luo}, {Sun}, {Xue}, {Bauer}, {Koekemoer}, {Lehmer}, {Liu}, {Schneider},
  {Shemmer}, {Trump}, {Vignali}, \& {Wang}}]{yang_2017}
{Yang}, G., {Chen}, C. T.~J., {Vito}, F., {et~al.} 2017,
  \href{http://dx.doi.org/10.3847/1538-4357/aa7564}{\JournalTitle{\apj}, 842,
  72}

\end{thebibliography}
